\pdfoutput=1
\documentclass[amsmath,amssymb,aps,prl,showpacs,twocolumn,groupedaddress]{revtex4-1}

\usepackage{graphicx}
\usepackage{subfigure}
\usepackage[T1]{fontenc}
\usepackage{times}
\usepackage{mathtools}
\usepackage{tensor}
\usepackage{stmaryrd}
\usepackage{MnSymbol}

\usepackage{pdfpages}

\newcommand{\ve}[1]{\mathbf{#1}} % Vectors
\newcommand{\p}{\mathrlap{'}} % A prime that takes no space
\DeclareMathOperator{\Prob}{\mathbb{P}}

\newcommand{\system}{\mathcal{S}}
\newcommand{\systemsteady}{\widehat{\mathcal{S}}}
\newcommand{\empirical}{\mathcal{E}}
\newcommand{\mechanistic}{\mathcal{M}}
\newcommand{\mechanisticsteady}{\widehat{\mathcal{M}}}
\newcommand{\systemBTW}{\mathcal{S}_{\text{BTW}}}
\newcommand{\systemBTWsteady}{\widehat{\mathcal{S}}_{\text{BTW}}}
\newcommand{\systemBTWmu}{\mathcal{S}_{\text{BTW}}^\mu}
\newcommand{\systemBTWmusteady}{\widehat{\mathcal{S}}_{\text{BTW}}^{\,\mu}}
\newcommand{\empiricalBTW}{\mathcal{E}_{\text{BTW}}}

\newcommand{\intermediateBTW}{\mathcal{I}_{\text{BTW}}}
\newcommand{\intermediateBTWmu}{\mathcal{I}_{\text{BTW}}^\mu}
\newcommand{\mechanisticBTW}{\mathcal{M}_{\text{BTW}}}
\newcommand{\mechanisticBTWmu}{\mathcal{M}_{\text{BTW}}^\mu}
\newcommand{\mechanisticBTWmusteady}{\widehat{\mathcal{M}}_{\text{BTW}}^\mu}

\newcommand{\inAB}[1]{{#1}\mathrlap{\kern-0.05em\smash{\raisebox{-0.5ex}{\rotatebox{20}{\ensuremath{\scriptscriptstyle\rcurvearrowright}}}}}}
\newcommand{\funH}{H}
\newcommand{\funA}[2]{\tensor*[_{\inAB{#2}}^{#1}]{A}{}}
\newcommand{\funB}[2]{\tensor*[_{\inAB{#2}}^{#1}]{B}{}}
\newcommand{\funHxyz}{\funH(w,x,\ve{y},\ve{z})}
\newcommand{\funAxyz}[2]{\funA{#1}{#2}(w,x,\ve{y},\ve{z})}
\newcommand{\funBxyz}[2]{\funB{#1}{#2}(w,x,\ve{y},\ve{z})}

\begin{document}

\title{Controlling Self-Organizing Dynamics on Networks Using Models that Self-Organize}

\author{Pierre-Andr\'e No\"el}
\email{noel.pierre.andre@gmail.com}
\author{Charles D. Brummitt}
\affiliation{University of California, Davis, California 95616, USA}
\author{Raissa M. D'Souza}
\altaffiliation[Also at ]{The Santa Fe Institute, Santa Fe, New Mexico 87501, USA}
\affiliation{University of California, Davis, California 95616, USA}

\date{August 12th 2013}
%
% 89.75.Hc Networks and genealogical trees
% 02.30.Yy Control theory 
% 05.65.+b Self-organized systems
% 45.70.Ht Avalanches [within 45.70.-n Granular systems]
\pacs{89.75.Hc, 02.30.Yy, 05.65.+b, 45.70.Ht}
%
%****************************************************************
\begin{abstract}
Controlling self-organizing systems is challenging because the system responds to the controller. Here we develop a model that captures the essential self-organizing mechanisms of Bak-Tang-Wiesenfeld (BTW) sandpiles on networks, a self-organized critical (SOC) system. This model enables studying a simple control scheme that determines the frequency of cascades and that shapes systemic risk. We show that optimal strategies exist for generic cost functions and that controlling a subcritical system may drive it to criticality. This approach could enable controlling other self-organizing systems.
\end{abstract}
\maketitle
%
%****************************************************************
%
Complex, self-organizing systems are challenging to control because their feedback mechanisms make it difficult to predict the effects of perturbations. For example, strategies for vaccination and quarantine must account for the human--disease feedback, namely, that people's behavior affects the spread of epidemics and epidemics affect people's behavior~\cite{Perra2011}. Accounting for feedbacks is especially important for controlling systems poised near a critical point because small changes can cause dramatic consequences. Many engineered and natural systems---such as forest fires~\cite{Malamud1998}, power grids~\cite{Dobson2007}, water reservoirs~\cite{Mamede2012}, brains~\cite{Ribeiro2010,Haimovici2013}, economies~\cite{Bak1993,Norrelykke2002}, and financial markets~\cite{Dupoyet2011}---appear to self-organize toward critical points with power-law-distributed event sizes, a phenomenon called self-organized criticality (SOC). Thus, controlling these systems can profoundly affect systemic risk (i.e., the chance of system-wide catastrophe). For example, suppressing small blackouts in power grids may increase the risk of large ones~\cite{Dobson2007}; making grids ``smart'' by adding meters, controlling loads, and introducing differential pricing \cite{Moslehi2010,Blumsack2012} could affect reliability in diverse ways. To meet the challenge of controlling self-organizing systems, here we use analytical models that capture the system's feedback mechanisms, i.e., models that self-organize.

Consider a complex system $\system$ that self-organizes to a stationary state $\systemsteady$. To simplify and understand $\system$, we model $\system$ in one of two ways. An \emph{empirical} model $\empirical$ contains parameters measured from the stationary state $\systemsteady$. By contrast, a \emph{mechanistic} model $\mechanistic$ contains only mechanistic rules (without empirical measurements of $\systemsteady$). Although $\empirical$ provides insight into $\systemsteady$, $\empirical$ cannot predict the effects of controlling the system away from $\systemsteady$. However, if $\mechanistic$ self-organizes to a stationary state $\mechanisticsteady$ via mechanisms like those that drive $\system$ to $\systemsteady$, then controlling $\mechanistic$ to change $\mechanisticsteady$ can efficiently predict ways to control $\system$ to change $\systemsteady$, thus solving the open problems of reducing the systemic risk of and enhancing the function of self-organizing systems~\cite{Perra2011,Malamud1998,Dobson2007,Mamede2012,Ribeiro2010,Haimovici2013,Dupoyet2011}. 

In this Letter, we contribute a successful example of controlling an SOC system using a mechanistic model. Here, the system is the BTW sandpile process on a network, denoted by $\systemBTW$, and $\empiricalBTW$ is a well-studied past model~\cite{Goh2003}. Our mechanistic model $\mechanisticBTW$ is a multitype branching process that self-organizes by fixing its free parameters via self-consistency. Using $\mechanisticBTW$, we systematically evaluate a control scheme for $\systemBTW$, an exercise expensive to simulate with $\systemBTW$ and impossible with $\empiricalBTW$. Specifically, we control how often cascades occur, which affects how large cascades could plausibly be. The results illuminate the tradeoffs that plague strategies to control many natural, financial, and infrastructure systems: frequently triggering cascades mitigates large events but sacrifices short-term profit, while avoiding cascades maximizes short-term profit but suffers from rare, massive events. We expect self-organizing, multitype branching processes like $\mechanisticBTW$ to inform controlling other self-organizing systems, including multistate dynamics with information bouncing back-and-forth on networks.
%
%****************************************************************
\paragraph{BTW sandpile on a network.\rlap{---}}
The presence of power laws in the magnitudes of events occurring in many real-world systems is often attributed to SOC~\cite{Malamud1998,Dobson2007,Mamede2012,Ribeiro2010,Haimovici2013,Dupoyet2011,Bak1987,Bak1988}. Typically, two competing mechanisms dominate: large events slowly but steadily become more probable, whereas the probability of future large events decreases when a large event occurs. For example, tectonic energy builds and then releases in earthquakes~\cite{OFC_quenchednetwork}. As another example, investment managers or infrastructure stakeholders balance cost and fear: they may slowly increase risk for profit, but when catastrophe occurs they mitigate risk via self-moderation or imposed regulations.

The BTW sandpile process~\cite{Bak1987,Bak1988} is an archetypal example of such mechanisms. We slowly add grains of sand (interpreted as load) to the system, which increases the chance of large cascades, but grains dissipate (disappear) during cascades. Originally introduced on the 2D lattice~\cite{Bak1987,Bak1988}, the BTW sandpile process has since been generalized to networks in a few natural ways that differ only in specifics~\cite{sandpile_wattsstrogatz_2d,Hoore2013, sandpile_wattsstrogatz_1d, sandpile_ER_annealed,OFC_quenchednetwork,Goh2003,Lee2004,Goh2005,Lee2012,Brummitt2012}. In this Letter, we consider the following formulation~\cite{Goh2003,Lee2012,Brummitt2012}.

The system $\systemBTW$ consists of a network of $N$ nodes that hold grains of sand. The structure of the network is fixed, but the amount of sand on each node changes in time. We call a node \emph{$i$-sand} if it holds $i$ grains of sand. The \emph{capacity} of a node is the maximal amount of sand that it can hold. In this Letter, we set the capacity of every node to one less than its degree (number of neighbors)~\cite{sandpile_ER_annealed, Goh2003, Lee2004, Goh2005, Lee2012, Brummitt2012}. Hence, a $2$-sand node of degree $3$ is \emph{at capacity}, meaning that it holds as much sand as it can withstand. Adding a grain to this node brings it \emph{over capacity}. A node over capacity \emph{topples}, meaning that it sheds its load by sending one grain to each of its neighbors.

The process consists of \emph{cascades} (avalanches) defined as follows. Drop a grain of sand on a node chosen uniformly at random, called the \emph{root} of the cascade. If this addition does not bring the root over capacity, then that cascade is finished. Otherwise, the root \emph{topples} by shedding one grain to each of its neighbors. Any node that now exceeds its capacity topples in the same way, until all nodes are under or at capacity. Whenever a grain of sand moves from one node to another, it dissipates (disappears) independently with probability $\epsilon$. The \emph{size} of a cascade is the number of toppling events, while the \emph{area} of a cascade is the number of nodes that topple. We begin a new cascade by dropping a grain on a uniformly random root node. See the Supplemental Material (SM)~\cite{SM_BTW_PRL} for the algorithm used in simulations.

In the dual limit of infinite network size ($N \to \infty$) and then rare dissipation ($\epsilon \to 0$)~\footnote{Contrast (i) this dual limit $N \to \infty$, $\epsilon \to 0$ for the BTW process on a network with (ii) the single limit $N \to \infty$ for traditional formulations of the BTW process on a $2$-dimensional lattice. In (ii), sand dissipates upon being shed off the boundary of the lattice, and the relative size of the boundary to the whole lattice vanishes in the limit $N \to \infty$. Hence, the single limit in (ii) implicitly achieves vanishing dissipation comparable to the limit $\epsilon \to 0$ in (i).}, it has been shown that the system self-organizes to a critical state $\systemBTWsteady$ \cite{sandpile_ER_annealed,Goh2003}. Both the cascade area and the cascade size distribution then exhibit a power law with exponent $-\tau$, where $\tau = 3/2$ for random graphs with light-tailed degree distributions (the mean-field case) and $\tau = \gamma/(\gamma-1)$ for random graphs with power-law degree distributions of exponent $\gamma$~\cite{Goh2003}.

For simplicity, this Letter considers the BTW process on a random $3$-regular graph (i.e., a random network of degree-$3$ nodes). We define $\psi_0$, $\psi_1$, and $\psi_2$ to be the probabilities that a uniformly random node is $0$-, $1$-, or $2$-sand, respectively. Similarly, for all $i,j \in \{ 0, 1, 2 \}$, we define $ \phi_{ij}$ to be the probability of reaching a $j$-sand node by following a link from a uniformly random $i$-sand node. Note that the methods used in this Letter can generalize to networks other than $3$-regular, which will be considered in an upcoming publication.
%
%****************************************************************
\paragraph{Empirical and self-organizing models.\rlap{---}}
A simple, empirical model $\empiricalBTW$ of the BTW sandpile process on a random $3$-regular graph may be derived as follows. Assuming $N \to \infty$, the probability generating function (PGF) for cascade area, $G(x) \equiv \sum_{a = 0}^\infty \Prob(\text{area} = a) x^a$, can be obtained by a standard single-type branching process
\begin{subequations}%
\begin{align}%
  F(x) & = 1 - (1-\epsilon) \phi_{22} + (1-\epsilon) \phi_{22} x \bigl[F(x)\bigr]^2 \label{eq:empirical:alongedge}\\
  G(x) & = 1 - \psi_2 + \psi_2 x \bigl[ F(x) \bigr]^3 . \label{eq:empirical:fromroot}
\end{align}\label{eq:empirical}% 
\end{subequations}%
Equation~\eqref{eq:empirical} uses the empirical observation that in treelike graphs only nodes initially at capacity (i.e., nodes at capacity just before the cascade begins) can topple during this cascade; we rigorously prove this observation using a mechanistic perspective in the SM~\cite{SM_BTW_PRL}. The PGF $F(x)$ gives the contribution to the area of a node sending a grain to a neighbor $v$ that has not yet toppled: the grain reaches $v$ with probability $1-\epsilon$, and $v$ is at capacity with probability $\phi_{22}$. If both these events occur, then $v$ topples (factor $x$) and sends grains toward its $3$ neighbors, $2$ of which have not yet toppled (factor $[F(x)]^2$). In $G(x)$, the root is initially at capacity with probability $\psi_2$, in which case it topples (factor $x$) and sends a grain toward its $3$ neighbors (factor $[F(x)]^3$).

For fixed dissipation $\epsilon > 0$, we can measure $\psi_2$ and $\phi_{22}$ in simulations (i.e., in $\systemBTWsteady$) and use $\empiricalBTW$ [Eq.~\eqref{eq:empirical}] to approximate the probability distribution of cascade area. However, this empirical model cannot \emph{predict} the parameters $\psi_2$ and $\phi_{22}$ on its own. This lack of closure becomes more problematic if we control the system away from its ``natural'' observed state $\systemBTWsteady$. A simple mechanistic argument partially solves this problem: with $\langle s \rangle$ denoting the average cascade size, the balance of sand input and average dissipation requires $3 \epsilon \langle s \rangle = 1$ ($1$ grain added per cascade; $3$ grains shed by toppling; shed grains dissipate with probability $\epsilon$). Assuming finite-size cascades for $\epsilon > 0$ [i.e., $(1-\epsilon)\phi_{22} < 1/2$], the expected cascade area satisfies
\begin{align}
  \langle a \rangle & \equiv \sum_{a=0}^\infty a \Prob(\text{area} = a) = G'(1) = \frac{\psi_2 \bigl[ 1 + (1-\epsilon) \phi_{22} \bigr]}{1 - 2 (1-\epsilon) \phi_{22}} , \label{eq:expectedarea}
\end{align}
and we know from empirical observations that $\langle s \rangle \approx \langle a \rangle$. Hence, the criterion for balancing sand, $3 \epsilon \langle s \rangle = 1$, converts the empirical model $\empiricalBTW$ into an \emph{intermediate} model $\intermediateBTW$, where mechanistic arguments fix $\phi_{22}$ in terms of $\psi_2$. Next we use $\intermediateBTW$ to explore controlling the system $\systemBTW$. Later we derive a fully mechanistic, self-organizing model $\mechanisticBTW$ [Eq.~\eqref{eq:H}] that fixes all unknown parameters.
%
%****************************************************************
\paragraph{Control and cost.\rlap{---}}
\begin{figure}
\begin{center}
\includegraphics{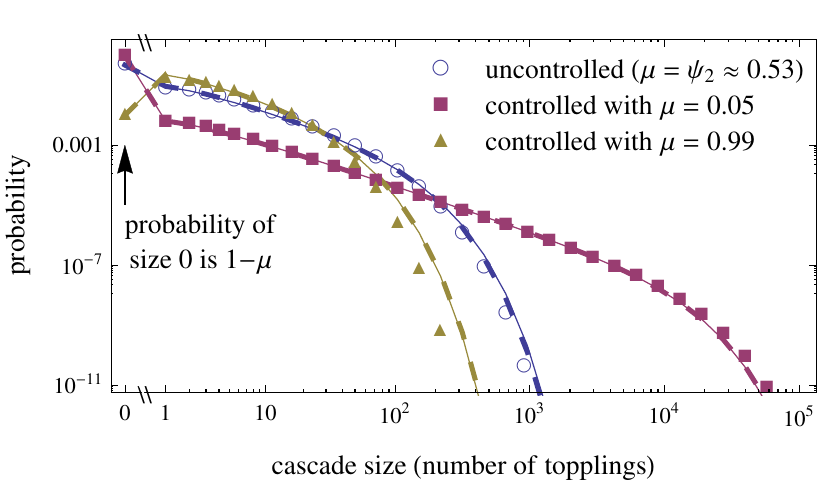}
\caption{(color online) Controlling the frequency of cascades $\mu$ significantly affects the cascade size distribution. The chance of no cascade (i.e., a size-$0$ cascade) is $1-\mu$, while the chance of a cascade of size $\geq 1$ is the control parameter $\mu$. In the original BTW model, $\mu$ is set to $\psi_2$, the fraction of $2$-sand (at capacity) nodes. Symbols denote results of simulations on random $3$-regular graphs with $\epsilon = 0.05, N=10^6$, while dashed and plain lines show the predictions of the intermediate model $\intermediateBTWmu$ and of the self-organizing model $\mechanisticBTWmu$, respectively.
\label{fig:control}}
\end{center}
\end{figure}
\begin{figure}
\begin{center}
\includegraphics{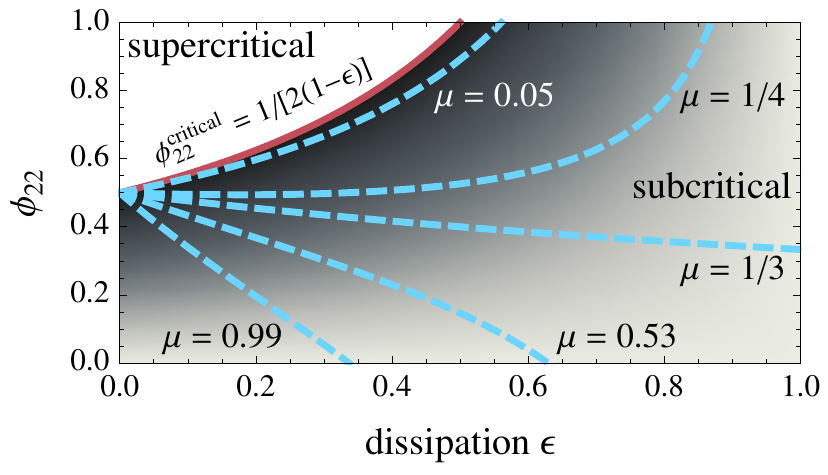}
\caption{(color online) Phase diagram of the controlled system $\intermediateBTWmu$, an approximation of $\systemBTWmu$ (similar to the diagram in Ref.~\cite{Lauritsen1996} except with control). Dashed lines are the system's attractors $\phi_{22} = (1-3 \epsilon \mu)/[(1-\epsilon) (3\epsilon \mu + 2)]$. For $\mu>0$, the system is critical only when $\epsilon \to 0$, but as $\mu \to 0$ the system approaches the critical line $\phi_{22}^{\text{critical}} = 1/[2(1-\epsilon)]$ for all $\epsilon < 1/2$. In the subcritical regime, darker shades denote proximity to criticality. \label{fig:phase}}
\end{center}
\end{figure}
Rather than suppressing sandpile cascades in just a specific region of a lattice~\cite{Cajueiro2010a,Cajueiro2010b} or steering the system to a particular state~\cite{Liu2011,CorneliusKathMotter2013}, here we control the stationary state $\systemBTWsteady$ of $\systemBTW$ to change the risk of small and large cascades. Our controller faces severe constraints: she can neither alter the value of $\epsilon>0$, nor the network, nor the cascade mechanism (unlike in Refs.~\cite{Cajueiro2010c,DAgostino2012}). Instead, the controller can only adjust where the first grain of sand of a cascade tends to land. Using some unspecified method, the controller sets the probability $\mu$ that the first grain lands on a $2$-sand node and hence causes a cascade (of size $\geq 1$). This rule defines the system $\systemBTWmu$, which reduces to the uncontrolled system, $\systemBTW$, when $\mu := \psi_2$. 

We obtain the \emph{controlled intermediate} model $\intermediateBTWmu$ by substituting $\mu$ for $\psi_2$ in $\intermediateBTW$ [i.e., Eqs.~\eqref{eq:empirical} and~\eqref{eq:expectedarea} and related], so $\phi_{22}$ is determined by $\epsilon$ and $\mu$. $\intermediateBTWmu$ can accurately predict the cascade size distribution as one varies $\mu$ (Fig.~\ref{fig:control}) without needing to observe $\phi_{22}$ empirically, but $\intermediateBTWmu$ cannot provide $\psi_2$, the fraction of $2$-sand nodes.

Increasing $\mu$ above $\psi_2$ is analogous to dropping snow where avalanches are about to occur and to starting forest fires in fire-prone areas, with the hope of preventing large avalanches and large fires in the long run. This strategy of triggering cascades $\mu = 99\%$ of the time (filled triangles of Fig.~\ref{fig:control}) does mitigate large avalanches, at the cost of causing more small ones (compared to the uncontrolled case, open circles). The other control strategy, decreasing $\mu$ below $\psi_2$, is akin to avoiding cascades as much as possible. Examples include extinguishing all forest fires or engineering power grids to suppress all blackouts, including small ones~\cite{Dobson2007}. This strategy (filled squares of Fig.~\ref{fig:control}) reduces the frequency of cascades to $\mu = 5\%$, at the cost of making the tail heavier.

The phase diagram of $\intermediateBTWmu$ (Fig.~\ref{fig:phase}) illustrates the essential behavior: controlling a subcritical system can make it critical. For fixed dissipation $\epsilon$ and control parameter $\mu$, the probability $\phi_{22}$ reaches a steady state (dashed lines). As $\epsilon \to 0$, the steady state collapses to $\phi_{22} = 1/2$ for all fixed $\mu \in (0,1]$ and for the uncontrolled system ($\mu := \psi_2$). However, for $\epsilon>0$, decreasing $\mu$ brings the system closer to criticality (darker shade of background in Fig.~\ref{fig:phase}) and reaches criticality when $\mu \to 0$. Thus, decreasing $\mu$ to avoid cascades leads to criticality and hence power-law-distributed event sizes (Fig.~\ref{fig:control}, squares). By contrast, increasing $\mu$ pushes the stationary state away from the critical line, hence mitigating large cascades (curve $\mu = 0.99$ in Fig.~\ref{fig:phase} and triangles in Fig.~\ref{fig:control}).

Both $\mu>\psi_2$ and $\mu<\psi_2$ have tradeoffs, so under what conditions is one strategy better? Because sand input equals average dissipation ($1 = 3 \epsilon \langle s \rangle$), we cannot control average cascade size $\langle s \rangle$ using the control parameter $\mu$. However, the cost of a cascade may grow nonlinearly with cascade size, in which case the average cost depends on $\mu$.

\begin{figure}
\begin{center}
\includegraphics{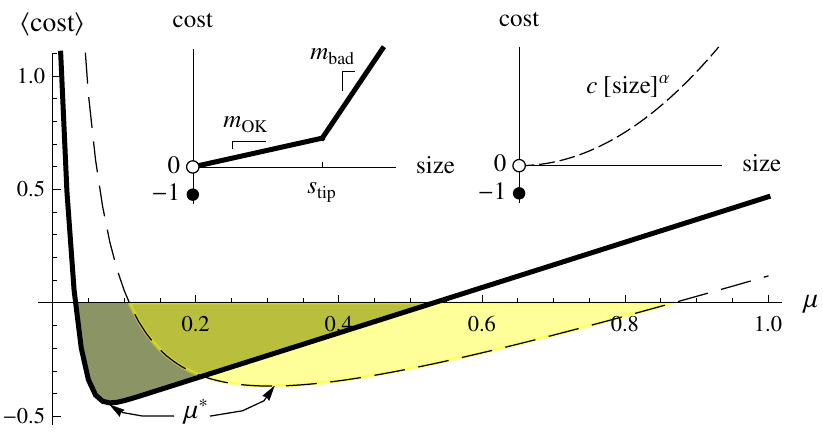}
\caption{(color online) If size-$0$ cascades confer benefit $1$ and if, as justified in the text, costs increase nonlinearly, such as with tipping points (left inset, thick lines) or as cascade size raised to a power $\alpha > 1$ (right inset, dashed lines), then there may exist a nontrivial, optimal control parameter $\mu^*$ that minimizes the expected cost in the stationary state $\systemBTWmusteady$ of the controlled SOC system $\systemBTWmu$. (Here, $\epsilon = 0.05$ and $m_\text{OK} = 0.07, m_\text{bad} = 0.5, s_\text{tip} = 10^4$; $c = 0.005, \alpha = 1.5$.)\label{fig:cost}}
\end{center}
\end{figure}

Here, we consider two concave cost functions illustrated in Fig.~\ref{fig:cost} (inset). First, motivated by the idea that small catastrophes in infrastructure are inexpensive to handle but that large disasters become expensive, we define a cost function with slope $m_\text{OK}$ for events smaller than a tipping point $s_\text{tip}$ and steeper slope $m_\text{bad}$ for events larger than $s_\text{tip}$. Our other cost function grows smoothly as the cascade size raised to a power $\alpha > 1$. [Both cost functions could arise from risk aversion (extra disutility to bad outcomes)~\cite{Newman2002_COLD}, government penalties for starting cascading failures, herdlike loss of consumer confidence, and/or indirect costs of disasters due to interdependencies with human health and with other infrastructures.] Finally, both cost functions assign a benefit of $1$ for size-$0$ cascades (in which no nodes topple); this benefit defines the scale of costs. (Infrastructures and investment portfolios, for instance, typically profit on uneventful days, yet catastrophes incur costs.)

For many parameters, these two concave cost functions have a nontrivial, optimal control parameter $\mu^*$ that minimizes the expected cost of cascades in the stationary state $\systemBTWmusteady$ (Fig.~\ref{fig:cost}). Increasing $\mu$ above $\mu^*$ mitigates large cascades but exacerbates small ones that accrue costs, while decreasing $\mu$ below $\mu^*$ makes cascades more rare but enables especially costly, massive cascades. The SM shows evidence that optimal $\mu^*$ is generic for heavy-tailed event distributions~\cite{SM_BTW_PRL}.

Finding or avoiding $2$-sand nodes becomes difficult when they become rare or widespread, respectively. To model this phenomenon, the controller could use her budget to apply a force $f$ to achieve a $\mu$ given by, e.g., $\mu(\psi_2,f) := \tan ^{-1} \pmb{(} f-\cot (\pi  \psi_2) \pmb{)} / \pi+1/2$, so that $f=0$ recovers $\systemBTW$, and pushing $\mu$ to $1$ or $0$ requires infinite force $f$. However, $\intermediateBTWmu$ does not provide $\psi_2$, so a closed model is needed.
%
%****************************************************************
\paragraph{Self-organizing model.\rlap{---}}
We now introduce a mechanistic, multitype branching process $\mechanisticBTWmu$ that self-organizes to a stationary state $\mechanisticBTWmusteady$ without using empirical measurements of $\systemBTWmusteady$. PGFs predict cascade outcome: $w$ and $x$ generate the cascade size and area, respectively, while $y_i$ and $z_i$ (elements of the vectors $\ve{y}$ and $\ve{z}$) generate the changes in the numbers of $i$-sand nodes and of $ii$-sand links (edges between two $i$-sand nodes), respectively. Unlike $\intermediateBTWmu$, self-consistency here fixes \emph{all} parameters $\psi_i$ and $\phi_{ij}$: on average, the numbers of $i$-sand nodes and of $ii$-sand links do not change at the stationary state. 

Like $G(x)$ [Eq.~\eqref{eq:empirical:fromroot}], the PGF $\funHxyz$ tracks the contribution of the cascade's root
\begin{align}%
  & \funH = \frac{1-\mu}{1-\psi_2} \sum_{\mathclap{i=0}}^1 \psi_i \frac{y_{i+1}}{y_i} \left[ \sum_{j=0}^2 \phi_{ij} \frac{1 + \delta_{(i+1)j}(z_j-1)}{1 + \delta_{ij}(z_j-1)} \right]^3 \label{eq:H} \\
  & + \mu x \sum_{\mathclap{n=1}}^\infty w^n \! \sum_{i'=0}^2 \frac{y_{i\p}}{y_i} \binom{3}{i\p} \bigl[ \funAxyz{i\p}{n} \bigr]^{3-i\p} \bigl[ \funBxyz{i\p}{n} \bigr]^{i\p}. \nonumber
\end{align}%
If the root is initially $i$-sand with $i \in \{0,1\}$ [first line of Eq.~\eqref{eq:H}], then it becomes $(i+1)$-sand and does not topple. Thus, the network has one fewer $i$-sand node (factor $y_i^{-1}$) and one more $(i+1)$-sand node (factor $y_{i+1}$). Furthermore, each link between the root and a $j$-sand neighbor warrants a factor $z_j^{-1}$ (respectively\ $z_j$) if $j = i$ (respectively\ $j = i+1$) to account for the lost (respectively\ new) $jj$-sand link. Note that only dyadic correlations (i.e., $\phi_{ij}$) are considered.

If the root is initially $2$-sand [second line of Eq.~\eqref{eq:H}], then it topples $n \ge 1$ times (factor $xw^n$) and ends up $i'\!$-sand after the cascade (factor $y_{i'}/y_i$), where $n$ and $i'$ depend on the number of grains that the root receives from its neighbors. Hence, a multitype branching process is required to count back-and-forth exchanges. In general, for a parent node $u$ with child $v$, we define a ``type'' for each combination $(n,n',i')$ such that, in a particular cascade, $u$ sends (respectively\ receives) a total of $n \ge 1$ grains toward (respectively\ $n'$ grains from) $v$, and, after the cascade, $u$ has $i\p \in \{0,1,2\}$ grains. The recurrence equation Eq.~\eqref{eq:empirical:alongedge} becomes a system of equations for two families of functions, $\funA{i\p}{n}$ and $\funB{i\p}{n}$, corresponding to the cases $n'=n-1$ and $n'=n$, respectively, which are the only possibilities for treelike graphs (see the proof and full expressions in the SM~\cite{SM_BTW_PRL}).

\begin{figure}
\begin{center}
\includegraphics{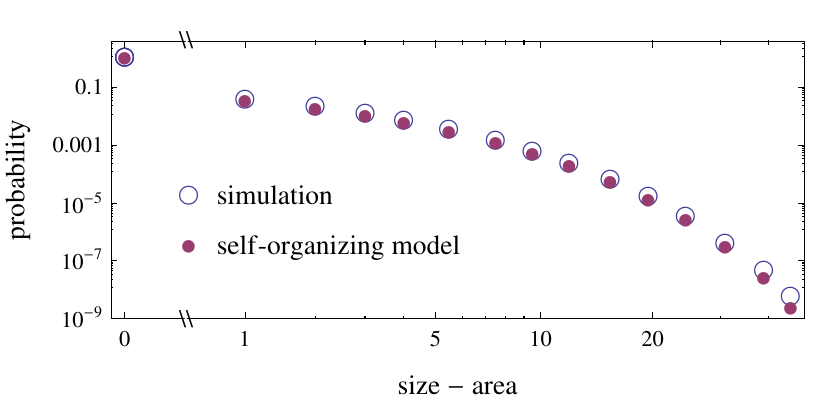}
\caption{(color online) Without any parameters from simulations, a self-organizing model can predict quantities inaccessible with past models, such as the area and the size of cascades. Here we plot the probability distribution for the difference between size and area for a random $3$-regular graph with $N=10^6, \epsilon = 0.001$.\label{fig:sizeminusarea}}
\end{center}
\end{figure}

Differentiating Eq.~\eqref{eq:H} with respect to $y_i$ and $z_i$ and setting all generators to $1$ gives $h_{(i)}$ and $\eta_{(i)}$, the average changes in the numbers of $i$-sand nodes and of $ii$-sand links. By hypothesis, the system has reached the stationary state, which provides the constraints $h_{(i)} = \eta_{(i)} = 0 \ \forall i \in \{ 0, 1, 2 \}$. Because the $\psi_i$ are probabilities and the $\phi_{ij}$ are conditional probabilities, they obey the additional constraints $\sum_{i = 0}^2 \psi_i = 1$, $\sum_{j = 0}^2 \phi_{ij} = 1 \ \forall i$, and $\psi_i \phi_{ij} = \psi_j \phi_{ji} \ \forall i,j$. Starting from educated guesses, numerical solution of the system of constraints provides values of $\psi_i$ and $\phi_{ij}$ consistent with those observed in the Monte Carlo simulations of $\systemBTWmu$, and it enables exploring ranges of parameters that would be computationally costly to simulate (large $N$, low $\epsilon$, and/or low $\mu$; see the SM~\cite{SM_BTW_PRL}). Values of $\psi_2$ obtained in this way may estimate the force $f$ required to achieve some control parameter $\mu(\psi_2,f)$. Finally, the PGF Eq.~\eqref{eq:H} distinguishes cascade size and area (Fig.~\ref{fig:sizeminusarea}), which to the best of our knowledge is a new result for BTW cascades on networks.

%
%****************************************************************
\paragraph{Future work.\rlap{---}}
Self-organizing branching processes could enable control of cascade area: if damage must occur, perhaps we can isolate it. For $\systemBTW$ with $\epsilon > 0$, a noninvasive control scheme cannot reduce the average cascade size because average sand input must be zero, but our mechanistic understanding of cascades could allow for $\langle \text{area} / \text{size} \rangle \approx 1/2$ in a treelike network and much smaller values in networks containing communities~\cite{SM_BTW_PRL}.

Adjusting the time scales so that more control occurs between cascades would make this model a dynamic version of Highly Optimized Tolerance (HOT)~\cite{Carlson2000} but with repeated cascades and control. Tuning the time scale between control and cascades could capture systems ranging from finance and brains (frequent cascades) to power grids and forest fires (infrequent cascades). 
%
%****************************************************************
\begin{acknowledgments}
The authors thank Kwang-Il Goh for useful discussion. This work was supported in part by the Defense Threat Reduction Agency Basic Grant No. HDTRA1-10-1-0088; the Army Research Laboratory Cooperative Agreement W911NF-09-2-0053; the Department of Defense (DoD) (C. D. B.); and the Fonds de recherche du Qu\'ebec--Nature et technologies (FRQNT) (P.-A. N.).
\end{acknowledgments}
%
%****************************************************************
%\bibliography{../btw_network_bibliography}
%****************************************************************
%Merlin.mbs v4.21 2009-07-09.
%
%
%****************************************************************
% Attach the Supplemental Material
%
\onecolumngrid\pagestyle{empty}


\section*{Supplementary material}

# Controlling self-organizing dynamics on networks using models that self-organize
## Supplemental Material (SM)

Pierre-André Noël, Charles D. Brummitt, and Raissa M. D'Souza[*]

*University of California, Davis, CA 95616*
(Dated: May 8, 2013)

This document provides supplemental material to the Letter "Controlling self-organizing dynamics on networks using models that self-organize" by the same authors. Section SM1 outlines the algorithm used in the numerical simulations. Section SM2 provides evidence that the behavior of the average cost of cascades for the controlled BTW system $\mathcal{S}^{\mu}_{\text{BTW}}$ is rather generic for heavy-tailed event size distributions. In Sec. SM3, we identify and prove results about the microscopic constraints on sandpile cascades. These constraints enable the self-organzing multitype branching process $\mathcal{M}^{\mu}_{\text{BTW}}$ given in full detail in Sec. SM4. Next, Sec. SM5 explicitly evaluates the derivatives of the generating function of the mechanistic model $\mathcal{M}^{\mu}_{\text{BTW}}$. Setting these derivatives to 0 in Sec. SM6 provides the self-consistency constraints, which is what makes this model $\mathcal{M}^{\mu}_{\text{BTW}}$ self-organize. Section SM6 also compares the results of this calculation with simulations. Finally, Sec. SM7 substantiates a conjecture made in the main article that, even though average cascade size is constrained in the BTW process, one may be able to control cascade area, with potentially significant results if the network has low expansion or community structure.

### SM1. SIMULATION ALGORITHM

The algorithm takes the following inputs: *Network_specification*, *Initial_condition*, $\epsilon$, $\mu$, *Convergence_iterations*, *Inner_statistics_iterations*, and *Outer_statistics_iterations*. The special case "$\mu := \psi_2$" (uncontrolled system $\mathcal{S}_{\text{BTW}}$) is handled by the algorithm by fixing the parameter $\mu = -1$; otherwise, $\mu$ must respect $0 < \mu \leq 1$ (controlled system $\mathcal{S}^{\mu}_{\text{BTW}}$). The algorithm returns three types of statistics: *Within_cascade_stats* (e.g., number of grains that fall on a degree 3 node at capacity), *After_cascade_stats* (e.g., cascade size), and *Network_stats* (e.g., number of degree 3 nodes at capacity). From a high-level perspective, the algorithm proceeds as follows.

---

Create the network according to *Network_specification*.    ▷ For example, make a random 3-regular network of $N$ nodes.
Set the amount of sand on each nodes according to *Initial_condition*.    ▷ Not necessary, but accelerates convergence.
**for** $i_{\text{conv}} = 1$ **to** *Convergence_iterations* **do**    ▷ Perform a great number of cascades to reach the steady state.
    Perform one cascade.    ▷ Note that no statistics are accumulated here.
**end for**
Initialize *Within_cascade_stats*, *After_cascade_stats*, and *Network_stats*.
**for** $i_{\text{outer}} = 1$ **to** *Outer_statistics_iterations* **do**
    **for** $i_{\text{inner}} = 1$ **to** *Inner_statistics_iterations* **do**
        Perform one cascade while updating *Within_cascade_stats*.  ▷ See next block of pseudocode at the end of this section.
        Update *After_cascade_stats*.    ▷ Updated *Outer_statistics_iterations* × *Inner_statistics_iterations* many times
    **end for**
    Update *Network_stats*.    ▷ These statistics concerning the whole network are updated *Outer_statistics_iterations* many times.
**end for**
**return** *Within_cascade_stats*, *After_cascade_stats*, and *Network_stats*.

---

A single run of the algorithm considers a single network structure. We verified that, in the large network limit, the outcomes varied little for different realizations of configuration models, such as the random 3-regular network considered in the main text.

---

[*] Also at The Santa Fe Institute, Santa Fe, NM 87501

Without detailing how the statistics are updated, we next give pseudocode for the part of the algorithm that performs one cascade.

---

Initialize *FIFO*, an (initially empty) "First In, First Out" queue.    ▷ Contains nodes that are scheduled to receive a grain.
*root* ← [a node selected uniformly at random].    ▷ Pick a first candidate as the root of the cascade.
**if** $0 < \mu \leq 1$ **then**    ▷ Do the following only in the controlled case $\mathcal{S}_{\text{BTW}}^{\mu}$.
    **if** $\mu >$ [random number from $[0, 1)$] **then**    ▷ Decide if the root should be at capacity.
        **while** *root* is not at capacity **do**    ▷ Retry until a root candidate at capacity is found.
            *root* ← [a node selected uniformly at random].    ▷ Pick a new root candidate.
        **end while**    ▷ Note that this is not the most efficient way to achieve this goal.
    **else**    ▷ The root should not be at capacity.
        **while** *root* is at capacity **do**    ▷ Retry until a root candidate not at capacity is found.
            *root* ← [a node selected uniformly at random].    ▷ Pick a new root candidate.
        **end while**
    **end if**
**end if**
Push *root* to the back of *FIFO*.    ▷ The root is now chosen, so schedule a grain to fall on it.
**while** *FIFO* is non-empty **do**    ▷ The cascade ends when all grains are resolved.
    *receiver* ← [the front element of *FIFO*].    ▷ Resolve the next grain, which will fall on the node with label *receiver*.
    Pop out the front element of *FIFO*.    ▷ The front element is now copied in *receiver*, thus no longer needed.
    Increment by one the number of grains on *receiver*.    ▷ Add the grain to *receiver*.
    **if** the number of grains on *receiver* is equal to *receiver*'s degree **then**    ▷ Does this bring *receiver* over its capacity?
        Set the number of grains on *receiver* to zero.    ▷ Yes, so *receiver* topples. Remove all grains from *receiver*.
        **for all** *neighbor* ∈ {neighbors of *receiver*} **do**    ▷ The grains removed from *receiver* are sent to its neighbors.
            **if** $(1 - \epsilon) >$ [random number from $[0, 1)$] **then**    ▷ Will this grain reach its target without dissipating?
                Push *neighbor* to the back of *FIFO*.    ▷ Yes, so schedule it to fall on *neighbor*.
            **end if**
        **end for**
    **end if**
**end while**
**return**

---

## SM2. COST ANALYSIS IN A GENERIC CONTROLLED-SOC CONTEXT

In the main text, we studied the average cost of cascades as a function of the control parameter $\mu$ for the BTW sandpile process $\mathcal{S}_{\text{BTW}}^{\mu}$. Notice that the cascade size distribution presented in Fig. 1 of the main text can be well approximated by a power law with exponential cutoff. This behavior is rather generic: power laws with exponential cutoff appear in the event size probability distributions of numerous engineered and natural systems (e.g., forest fire areas, solar flare intensities, earthquake magnitudes, website hits, phone calls, emails and more [1]). The reason why some of these systems exhibit such a distribution may be similar to what occurs in $\mathcal{S}_{\text{BTW}}$: there is a slow but steady increase in the risk of large events (new grain added at each cascade), and events mitigate the risk proportionally to their size (sand dissipation). Hence, it is possible that such systems behave similarly to $\mathcal{S}_{\text{BTW}}^{\mu}$ under control.

Here we repeat and expand the cost-analysis of the main text, but this time we assume that the cascade-size distribution is *exactly* a power law distribution with exponential decay. This assumption is somewhat accurate for $\mathcal{S}_{\text{BTW}}^{\mu}$ and plausible for many kinds of systems with event size distributions approximately described by a power law with exponential decay [1]. Using the same cost functions as in the main text, we observe that the average cost as a function of $\mu$ is qualitatively unaffected by the exact shape of the event size distribution. This qualitative agreement reinforces the hypothesis that engineered and natural systems with a similar self-organization mechanism could exhibit similar optima in their cost analysis.

We thus assume that the cascade size distribution $\mathbb{P}(\text{size} = s)$ of $\mathcal{S}_{\text{BTW}}^{\mu}$ has the form

$$\mathbb{P}(\text{size} = s) \equiv (1-\mu)\delta_{s0} + \mu(1-\delta_{s0})\frac{t^{-3/2} e^{-s/s_c}}{\text{Li}_{3/2}\left(e^{-1/s_c}\right)}, \tag{SM1}$$

where $\delta_{s0}$ is Kronecker's delta, $\text{Li}_\tau(z) \equiv \sum_{k=1}^{\infty} z^k/k^\tau$ is the polylogarithm, $s_c$ is an exponential cutoff (fixed by the balancing sand requirement, discussed next), and the power law exponent $-3/2$ is the mean-field exponent for the sandpile process [2].

(The precise value of the power law exponent does not affect the qualitative behavior of the average cost as a function of $\mu$ [Fig. 2(a)].) Recall that, in the equilibrium state of $\mathcal{S}_{\text{BTW}}^{\mu}$, sand input (of 1 per cascade) equals average sand dissipation (of $3\epsilon\mu\langle s\rangle$ per cascade for a random 3-regular graph). Given $\mu$ and $\epsilon$, this requirement that $1 = 3\epsilon\mu\langle s\rangle$ determines the cutoff $s_c$ of Eq. (SM1). By computing the average of the cascade size distribution (SM1), balancing sand in the system requires

$$\frac{1}{3\epsilon\mu} = \frac{\text{Li}_{1/2}(e^{-1/s_c})}{\text{Li}_{3/2}(e^{-1/s_c})}. \tag{SM2}$$

Equation (SM2) can be solved numerically to fix $s_c$ in terms of $\epsilon$ and $\mu$.

Next we compute the expected cost of cascades for the two cost functions defined in the main text and illustrated in the inset of Fig. 3 (and illustrated again in the inset of Fig. SM1). For the cost function $C_{\text{tip}}(s)$ that has benefit 1 (i.e., cost $-1$) for size $s = 0$, slope $m_{\text{OK}}$ for cascade sizes $s$ between 1 and a tipping point $s_{\text{tip}}$ and then steeper slope $m_{\text{bad}}$ above $s_{\text{tip}}$,

$$C_{\text{tip}}(s) \equiv \begin{cases} -1 & \text{if } s = 0 \\ m_{\text{OK}} s & \text{if } 1 \leq s \leq s_{\text{tip}} \\ m_{\text{bad}}(s - s_{\text{tip}}) + m_{\text{OK}} s_{\text{tip}} & \text{if } s > s_{\text{tip}} \end{cases}, \tag{SM3}$$

the expected cost is

$$\langle C_{\text{tip}}\rangle \equiv \sum_{s=0}^{\infty} \mathbb{P}(\text{size} = s) C_{\text{tip}}(s)$$
$$= b(\mu - 1) + \frac{\mu \exp\left(-\frac{s_{\text{tip}}+1}{s_c}\right)}{\text{Li}_{\frac{3}{2}}\left(\exp\left(-\frac{1}{s_c}\right)\right)} \Bigg[ s_{\text{tip}}(m_{\text{OK}} - m_{\text{bad}}) \Phi\left(\exp\left(-\frac{1}{s_c}\right), \frac{3}{2}, s_{\text{tip}} + 1\right)$$
$$+ (m_{\text{bad}} - m_{\text{OK}}) \Phi\left(\exp\left(-\frac{1}{s_c}\right), \frac{1}{2}, s_{\text{tip}} + 1\right) + m_{\text{OK}} \text{Li}_{\frac{1}{2}}\left(\exp\left(-\frac{1}{s_c}\right)\right) \exp\left(\frac{s_{\text{tip}}+1}{s_c}\right) \Bigg],$$

where $\Phi(z, s, a) \equiv \sum_{k=0}^{\infty} z^k (a+k)^{-s}$ is the Hurwitz–Lerch transcendent.

For the cost function $C_\alpha(s)$ that grows as a power $\alpha > 1$ with scale parameter $c > 0$,

$$C_\alpha(s) = \begin{cases} -1 & \text{if } s = 0 \\ cs^\alpha & \text{if } x \geq 1 \end{cases}, \tag{SM4}$$

the expected cost is

$$\langle C_\alpha\rangle \equiv \sum_{s=0}^{\infty} \mathbb{P}(\text{size} = s) C_\alpha(s) = \frac{\mu \text{Li}_{3/2-\alpha}(\exp(-1/s_c))}{\text{Li}_{3/2}(\exp(-1/s_c))} - (1 - \mu). \tag{SM5}$$

Figure SM1 plots the expected costs $\langle C_{\text{tip}}\rangle$ and $\langle C_\alpha\rangle$ as functions of the control parameter $\mu$, with the exponential cutoff of the size distribution fixed by the balancing sand requirement [Eq. (SM2)]. Figure SM1 is analogous to Fig. 3 but for the size distribution set to the ansatz given in Eq. (SM1). In Fig. SM1, we chose parameters of the cost functions so that the optimal value $\mu^*$ is smaller for the cost function $C_\alpha$ than for the cost function $C_{\text{tip}}$, which is the opposite of the order of the minima in Fig. 1 of the main text. (Note that the difference in the order of the minima in Fig. SM1 is not due to the cascade size distribution being Eq. (SM1) rather than simulations of the sandpile process $\mathcal{S}_{\text{BTW}}^{\mu}$ but instead due to changes in the parameters of the cost functions (SM3) and (SM4) in Fig. SM1 compared to the parameters used in Fig. 3.)

Because the particular shape of the expected cost versus $\mu$ curve (Fig. SM1) and the value $\mu^*$ of this curve's minimum depends on the parameters of the cost functions, we next explore how these cost curves and their minima change as we vary the shape parameter $\alpha \in [1, 2]$ of the cost function $C_\alpha$ [Eq. (SM4)]. Figure 2(a) shows that increasing $\alpha$ from 1 to 2 [i.e., changing from bottom to top curves in Fig. 2(a)] penalizes large cascades more and more, so the optimal $\mu^*$ increases in order to trigger frequent cascades to prevent massive ones. Conversely, decreasing $\alpha$ puts more emphasis on the costs of small cascades, so the optimal control parameter $\mu^*$ decreases in order to prevent cascades from occurring. The disks in Fig. 2(b) shows how the location of the optimal control parameter $\mu^*$ changes with $\alpha \in [1, 2]$, the extent to which large cascades are especially costly.

Decreasing the dissipation rate $\epsilon$ extends the tail of the cascade size distribution [via the sand balance requirement, Eq. (SM2)]. Thus, the optimal $\mu^*$ increases more quickly as large cascades become increasingly penalized (i.e., as $\alpha$ increases), as illustrated by changing from squares to circles to triangles in Fig. 2(b). Furthermore, decreasing $\epsilon$ and increasing $\alpha$ makes massive events more likely and more costly, respectively, which explains the large, red triangles in Fig. 2(b) (the areas of which depict the large expected costs). Thus, Fig. 2(b) illustrates the limitations of the $\mu$-control strategy: because $\mu$ cannot exceed 1, the control strategy cannot avoid large cascades to an arbitrary extent, so dangerous systems (with small $\epsilon$ and large $\alpha$) can be costly.

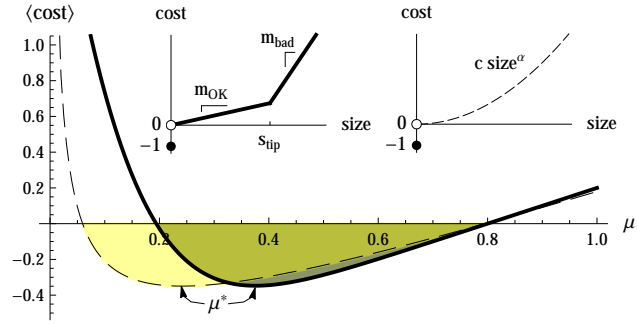

FIG. SM1. For the cascade size distribution given by a power law with exponential decay [Eq. (SM1)], the expected costs $\langle C_{\text{tip}}\rangle$ and $\langle C_\alpha \rangle$ (plotted in thick and dashed lines, respectively, and depicted in the insets) have nontrivial (i.e., $\neq 0, 1$) minima at $\mu^* \approx 0.375 \pm 0.001$ and $\mu^* \approx 0.241 \pm 0.001$, respectively. Here, the cost functions have the following parameters: the dissipation rate $\epsilon = 0.05$ for both curves, and $m_{\text{OK}} = .03, m_{\text{bad}} = .5, s_{\text{tip}} = 10^3$; $c = 0.01, \alpha = 1.3$. Note that the order of the two $\mu^*$ values is the reverse of their order in Fig. 1.

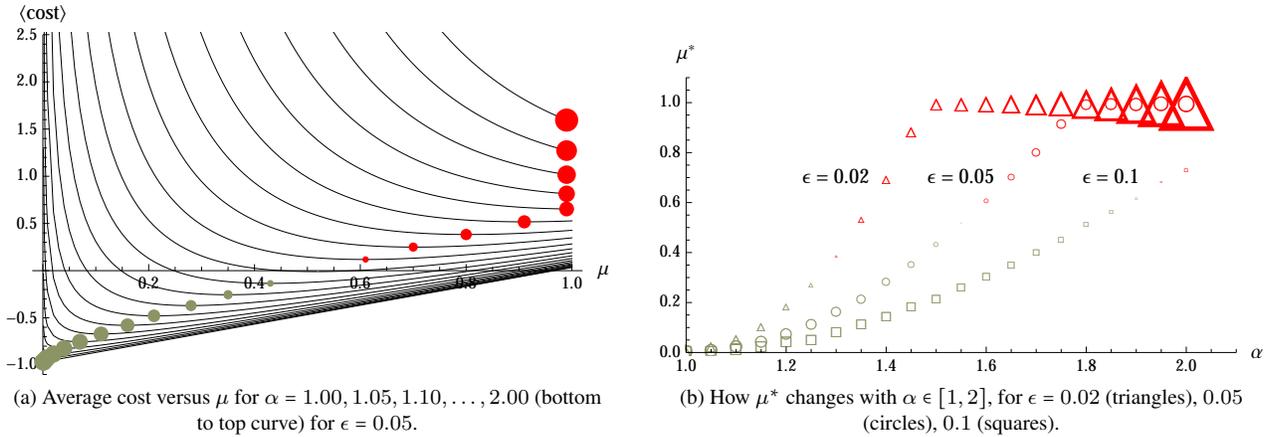

(a) Average cost versus $\mu$ for $\alpha = 1.00, 1.05, 1.10, \ldots, 2.00$ (bottom to top curve) for $\epsilon = 0.05$.

(b) How $\mu^*$ changes with $\alpha \in [1, 2]$, for $\epsilon = 0.02$ (triangles), 0.05 (circles), 0.1 (squares).

FIG. SM2. To show how the optimal control parameter $\mu^*$ changes as the cost function changes, we vary the shape parameter $\alpha$ of the cost function $C_\alpha(s)$ with $c = 0.005$. [Recall from Eq. (SM4) that $C_\alpha(s) := (-1)\delta_{s0} + (1 - \delta_{s0})cs^\alpha$.] In panel (a), we plot the average cost $\langle C_\alpha \rangle$ versus $\mu$ using Eq. (SM5). The disks denote the optimal $\mu^*$ that minimizes the expected cost. In both panels (a) and (b), the area of each symbol is proportional to the expected cost, and the color is red for positive costs and dollar-bill-green for negative costs. In panel (b), we plot the optimal $\mu^*$ that minimizes the expected cost $\langle C_\alpha \rangle$ as a function of $\alpha \in [1, 2]$ for $\epsilon = .02$ (triangles), $\epsilon = 0.05$ (circles) and $\epsilon = 0.1$ (squares).

## SM3. MICROSCOPIC UNDERSTANDING OF THE SANDPILE MODEL ON NETWORKS

In tree-like sandpile cascades on (not necessarily tree-like) networks, causality constrains the shape of cascades, the number of times that each node topples, and how many grains neighboring nodes exchange. The purpose of this section is to identify such constraints because they are essential to our later calculations of cascade area and size. Understanding Corollary 1 and the associated Figs. SM3–SM4 suffices to understand the sections that follow, whereas Theorems 1–2 are the main mathematical results that concisely summarize our characterization of the sandpile model on networks. Because Corollary 1 follows from Theorems 1–2, its proof is deferred to the end of this section.

The results obtained here hold for any network $\mathcal{G}$ (i.e., not only for random 3-regular networks), although we obtain strongest results when the cascade forms a tree (which is the case, or nearly the case, for most cascades on tree-like networks, including random 3-regular networks). To precisely define this notion of a cascade forming a tree, for a cascade on the graph $\mathcal{G}$, we define the graph $\mathcal{G}^\dagger$ to have all the nodes of $\mathcal{G}$ that have sand sent toward them in the cascade and all the edges of $\mathcal{G}$ along which sand is sent in the cascade. That is, $\mathcal{G}^\dagger$ is the subgraph induced by the root, the nodes that topple, and the neighbors of the nodes that topple, from which we remove the links between pairs of nodes that both do not topple. We say that *a cascade forms a finite tree* $\mathcal{G}^\dagger$ if $\mathcal{G}^\dagger$ is a finite tree.

For a cascade that forms a finite tree $\mathcal{G}^\dagger$, we associate a *pattern* to each node $v$ in $\mathcal{G}^\dagger$. A pattern inherits all properties of its associated node, such as whether the node is the root, whether it is at capacity, and whether it is a parent or child with respect to another node.

Each non-root pattern has a *signature* given by a pair of integers $(n, n')$, which characterizes grains exchanged between this pattern and its parent. Specifically, given a node $v$ with parent $u$, node $v$ is associated with a pattern of signature $(n, n')$ if and only if the parent $u$ sends $n$ grains toward $v$ and the parent $u$ receives $n'$ grains from $v$. Note that, due to dissipation, the child $v$ may receive fewer than $n$ grains from $u$, and $v$ may send more than $n'$ grains toward $u$. The intuition to keep in mind is that we count grains *from the parent's perspective* (i.e., $n, n'$ are the numbers of grains sent from and received by the parent $u$ with respect to this particular child $v$). Though each pattern has a single, well-defined signature, two different patterns may share the same signature. For simplicity, we say that a non-root node $v$ has signature $(n, n')$ if $v$ is associated to a pattern that has signature $(n, n')$.

The following corollary of Theorems 1–2 enumerates the rules (illustrated in Figs. SM3–SM4) that enable our mechanistic, self-organizing model $\mathcal{M}_{\text{BTW}}$ of the BTW process on networks.

**Corollary 1** (Constraints for patterns). *The following statements hold for a cascade that forms a finite tree $\mathcal{G}^\dagger$.*

*A node $v$ in $\mathcal{G}^\dagger$ topples $0$ times if and only if exactly one of the following holds:*

*(i). $v$ is the root and is not initially at capacity.*

*(ii). $v$ is non-root; $v$ is not initially at capacity; $v$ has signature $(1, 0)$; and the grain sent by $v$'s parent toward $v$ reaches $v$.*

*(iii). $v$ is non-root; $v$ is not initially at capacity; $v$ has signature $(1, 0)$; and the grain sent by $v$'s parent toward $v$ dissipates.*

*(iv). $v$ is non-root; $v$ is initially at capacity; $v$ has signature $(1, 0)$; and the grain sent by $v$'s parent toward $v$ dissipates.*

*For a positive integer $n$, a node $v$ in $\mathcal{G}^\dagger$ topples $n$ times if and only if exactly one of the following holds:*

*(v). $v$ is the root, and each of $v$'s children has signature $(n, n)$ or $(n, n-1)$, except not all of its children may have signature $(n, n)$.*

*(vi). $v$ is non-root; $v$ has signature $(n+1, n)$; and each of $v$'s children has signature $(n, n)$ or $(n, n-1)$, except not all of its children may have signature $(n, n)$.*

*(vii). $v$ is non-root; $v$ has signature $(n, n)$; and each of $v$'s children has signature $(n, n)$ or $(n, n-1)$.*

*(viii). $v$ is non-root; $v$ has signature $(n+1, n)$; each of $v$'s children has signature $(n, n)$ or $(n, n-1)$; and the last grain sent by $v$'s parent toward $v$ dissipates.*

*(ix). $v$ is non-root; $v$ has signature $(n, n-1)$; each of $v$'s children has signature $(n, n)$ or $(n, n-1)$; and the last grain sent by $v$ toward its parent dissipates.*

The rest of this section formally proves these rules. Because intermediary results are required, we defer the proof of Corollary 1 to the end of this section. Readers interested only in the applications of these rigorous results can focus on Corollary 1 and on Figs. SM3–SM4.

Before proceeding with proofs, we note that characterizing sandpile cascades using causality constraints requires a notion of time. To make time considerations well defined and independent of the numerical implementation of the model, we assume that each grain sent from one node to another takes a positive and possibly random amount of time to reach its target or to dissipate, and we assume that nodes topple as soon as they exceed their capacity.

Note that the phrase "a node $v$ receives $n$ grains from a single neighbor" has the intended meaning "for at least one of $v$'s neighbors, $v$ received $n$ grains from that neighbor".

**Lemma 1** (Constraints on any cascade). *In any cascade, the following statements hold for any positive integer $n$:*

*(i). A non-root node cannot topple for the $n$th time before it receives $n$ grains from a single neighbor.*

*(ii). No node may receive $n$ grains from a single neighbor before the root topples an $n$th time.*

*Proof.* To prove (i), first consider the case that a non-root node $v$ has degree $k$, begins with $k-1$ grains before the cascade, and at some time $t$ in the cascade has received exactly $n-1$ grains from each of its $k$ neighbors. Then $v$ has $kn-1$ grains initially on it and sent to it by time $t$, so $v$ has toppled at most $n-1$ times by time $t$. By construction, $v$ must receive an $n$th grain from at least one neighbor before $v$ can topple an $n$th time. If $v$ received fewer than $n-1$ grains from one or more of its neighbors by time $t$, then $v$ still cannot topple an $n$th time until it receives an $n$th grain from a single neighbor. This argument proves (i).

To prove (ii), let $t$ be the first time in a cascade at which a non-root node receives an $n$th grain from a single neighbor. Let $v$ be one such node. Then a neighbor $u$ of $v$ toppled an $n$th time before time $t$. Suppose for contradiction that $u$ is not the root. In order to topple $n$ times before time $t$, node $u$ must have received at least $n$ grains from one of its neighbors by time $t$, which contradicts the definition of $t$. Thus $u$ must be the root, and claim (ii) follows. □

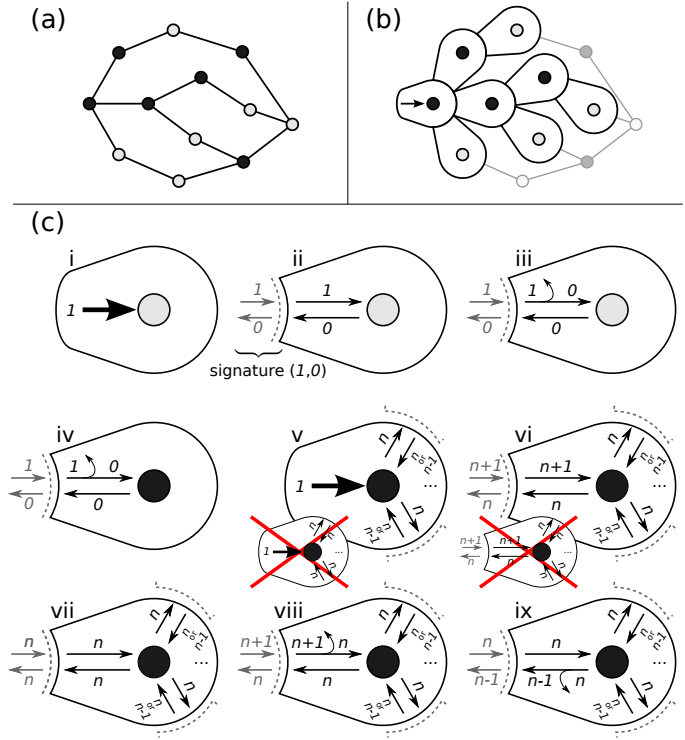

FIG. SM3. (Color online) Finite, tree-like cascades as an assemblage of patterns. (a) The graph before the cascade. Dark circles represent nodes at capacity, while light circles represent nodes not at capacity. (b) Starting at the root (indicated by an arrow), the cascade is assembled from patterns that encode how the grains of sand are shed during the process. Note that this assemblage may form a tree even though the original graph contained cycles. (c) The only possible patterns and rules for assembling them. The roman numerals correspond to the cases in Corollary 1. The convex shapes i and v represent root patterns, while the convex shapes represent non-root patterns. The large arrow represents the first grain dropped on the root; a curved arrow indicates dissipation; and the other arrows indicate exchanges of sand between neighboring nodes. The signature of a non-root pattern may be inferred by the arrows on its left. A large, red "X" mark a forbidden special case for patterns v and vi.

We are now ready to formulate strong constraints on the root (and on its surroundings) in any cascade.

**Theorem 1** (Strong constraints on the root in any cascade). *In any cascade, the root topples $0$ times if and only if the root was not initially at capacity. Moreover, for any positive integer $n$, the root topples $n$ times by time $t \geq 0$ if and only if (a) the root was initially at capacity, and (b) by time $t$ the root received from each of its neighbors either $n$ or $n-1$ grains (except not $n$ grains from each of its neighbors).*

*Proof.* To show the first claim, note that if the root is at capacity then it topples at least once, and if the root is not at capacity then it does not topple and the cascade ends there.

To prove the rest, fix $n \geq 1$. If (a) and (b) hold, then the number of grains initially on and received by the root (including the first grain dropped on it) by time $t$ is in the interval $[kn, k(n+1)-1]$, so the root topples $n$ times by time $t$.

Inversely, if (a) does not hold, then the root topples $0 < n$ times. It remains only to show that if (b) does not hold then the root does not topple $n$ times by time $t$. There are three cases. First, if the root has received $n$ grains from each of its neighbors by time $t$, then (by counting grains) we know that the root has toppled $n+1$ times by time $t$. Second, if the root has received $m > n$ many grains from a neighbor by time $t$, then that neighbor (call it $u$) must have toppled at least $m$ times by time $t$. Thus, $u$ must have received at least $m$ grains from at least one of its neighbors by time $t$. But by Lemma 1(ii), no node can receive $m$ grains from a single neighbor before the root topples $m$ times. Hence the root toppled at least $m > n$ times by time $t$.

In the third and final case, the root has received $m < n-1$ grains from at least one neighbor by time $t$. Let $t'$ be the time when the root topples for the $n$th time. Before $t'$, no neighbor of the root can have received $n$ grains from a single neighbor by Lemma 1(ii) because the root has toppled $\leq n-1$ times. Hence, by Lemma 1(i), no neighbor of the root can have toppled $n$ times before time $t'$. Thus, the root cannot have received $\geq n$ grains from a single neighbor by time $t'$. To conclude, the number of grains initially on and received by the root (including the first grain dropped on it) by time $t'$ is $\leq kn + 1 + (m-n) < kn$, which contradicts to the root toppling for the $n$th time at time $t'$. This concludes the proof. □

Using these results for any cascade, we now prove results for the case in which the cascade forms a tree.

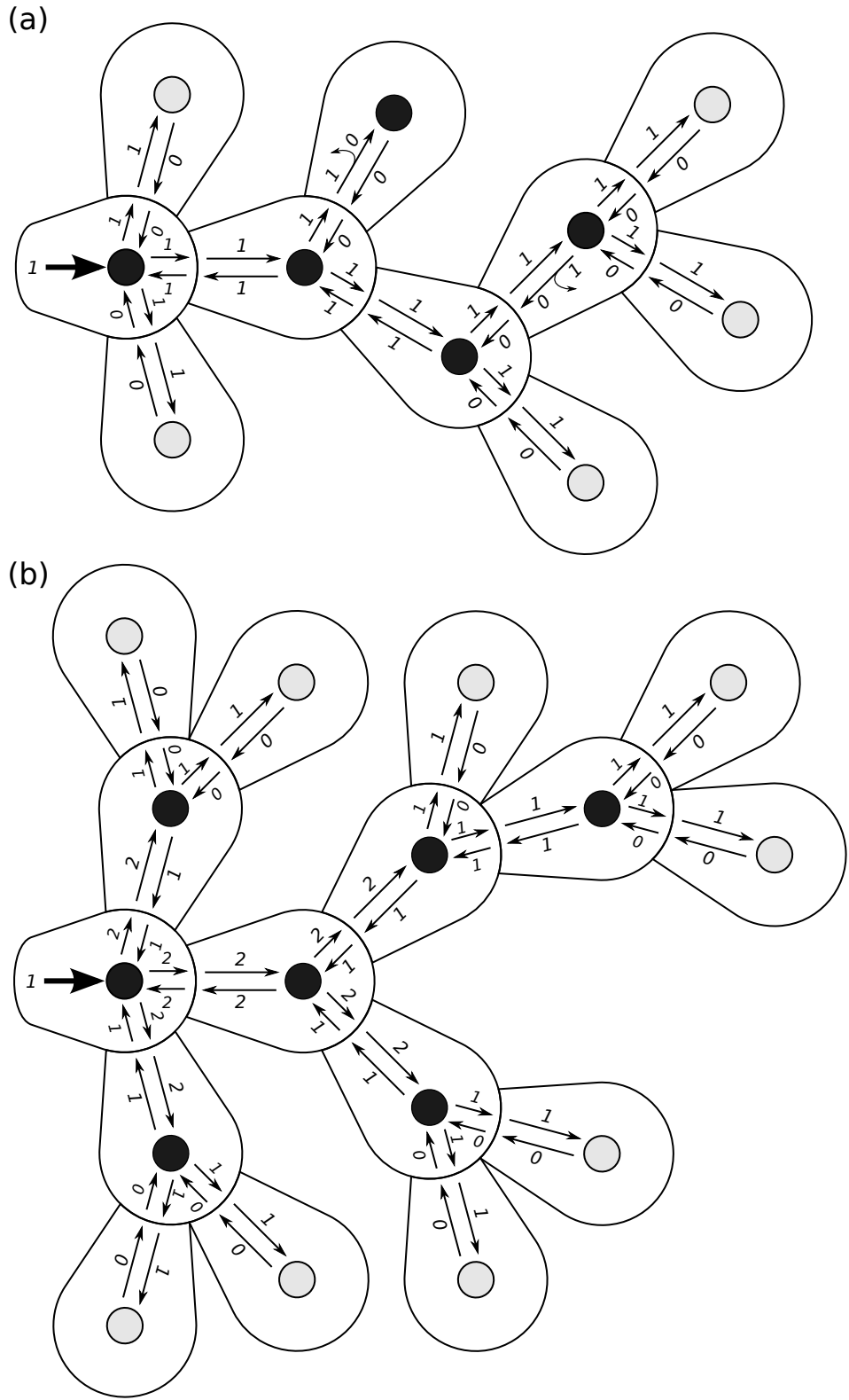

FIG. SM4. Two examples of tree-like cascades seen as valid assemblages of patterns. (a) There are 10 nodes in $\mathcal{G}'$, so 10 patterns are required. The root topples once and 3 non-root nodes topple once, so the cascade size and the cascade area are both 4. Five non-root nodes do not topple because they were not initially at capacity. A single non-root node initially at capacity does not topple because the grain sent toward it by its parent dissipates. A total of 2 grains dissipate in this cascade. (b) There are 16 nodes in $\mathcal{G}'$, so 16 patterns are required. The root topples twice, a single non-root node topples twice, and 5 non-root nodes topple once. Hence, the cascade size is 9 while the cascade area is 7. No grains dissipate in this cascade.

**Lemma 2** (Constraints on cascades that form a finite tree). *The following statements hold for a cascade that forms a finite tree $\mathcal{G}^\dagger$.*

  (i). *Let $n$ be a non-negative integer and $v$ be any node in $\mathcal{G}^\dagger$. No descendant of $v$ may receive an $n$th grain from a single neighbor before $v$ topples an $n$th time.*

 (ii). *Let $n$ be a non-negative integer and $v$ be any non-root node in $\mathcal{G}^\dagger$. Then node $v$ cannot topple an $n$th time before receiving an $n$th grain from its parent.*

(iii). *Let $n$ be a positive integer, and let $v$ be a non-root node that has received $n$ grains from its parent by time $t$. Then node $v$ has toppled $n$ times by time $t$ if and only if it is initially at capacity, and at the moment of toppling for the $n$th time it has received $n-1$ grains from every one of its children and $n$ grains from its parent.*

(iv). *Let $n$ be a positive integer, and let $v$ be a non-root node that has toppled $n$ times by time $t$. Then $v$'s parent toppled at most $n+1$ times by time $t$.*

 (v). *Let $v$ be any node in $\mathcal{G}^\dagger$. If $v$ is not at capacity, then $v$ topples $0$ times.*

*Proof.* To show (i), first note that if $v$ is the root, then the claim follows directly from the analogous (but weaker) result, Lemma 1(ii), for cascades that do not form trees. To finish proving (i), suppose $v$ is not the root, and assume (for contradiction) that a descendant of $v$ receives an $n$th grain from a single neighbor before $v$ topples an $n$th time. Let $t$ be the first time when a descendant of $v$ receives $n$ grains from a single neighbor; call such a descendant $u$, and let $w$ be a neighbor of $u$ from which $u$ receives an $n$th grain at time $t$. Then $w$ must have toppled for an $n$th time at time $t' < t$. We know that $w$ is not $v$ because $v$ does not topple for the $n$th time before time $t$. Because $w$ is a descendant of $v$, we know $w$ is not the root, so by Lemma 1(i) we know that $w$ must have received at least $n$ grains from a single neighbor before time $t'$, which contradicts the definition of $t$. Thus, claim (i) follows.

We show (ii) by contradiction. Suppose (for contradiction) that $v$ does not receive an $n$th grain from its parent by time $t$ and yet $v$ topples an $n$th time at time $t$. Before time $t$, $v$ has toppled fewer than $n$ times, so (by Lemma 2(i) applied to $v$) no children of $v$ have received $n$ grains from the same neighbor. Thus, no children of $v$ have toppled $n$ times by time $t$ [by Lemma 1(i)], so $v$ does not receive $n$ grains from the same child by time $t$. But by Lemma (i) and the assumption that $v$ topples at time $t$, we know $v$ must have received $n$ grains from a single neighbor before time $t$, a contradiction. Thus, claim (ii) follows.

Claim (iii) follows from counting grains of sand. Let $k$ be the degree of the non-root node $v$ that has received $n \geq 1$ grains from its parent by time $t$. Suppose that $v$ topples for an $n$th time at some time $t' \leq t$. Before time $t'$, Lemma 2(i) guarantees that the children of $v$ toppled at most $n-1$ times, so $v$ received at most $n-1$ grains from each of its children by time $t'$. Moreover, $v$ received at most $n$ grains from its parent by time $t'$ (because it received $n$ grains from its parent by time $t \geq t'$). Because $v$ topples for the $n$th time at time $t'$, the total number of grains initially on and received by $v$ by time $t'$ should be $kn$. From the preceding constraints, this is only possible if $v$ is initially at capacity, $v$ receives $n-1$ grains from each of its children by time $t'$, and $v$ receives $n$ grains from its parent by time $t'$.

Conversely, suppose that $v$ is initially at capacity, and suppose that $t'$ is the first time such that $v$ has received $n-1$ grains from every one of its children and $n$ grains from its parent. Then, by the previous grain-counting argument, $v$ topples an $n$th time at time $t'$. It remains to be proven that $v$ does not topple again by time $t$, which is guaranteed by Lemma 2(ii) because $v$ received only $n$ grains from its parent by time $t$. Thus claim (iii) holds.

We show (iv) by induction over the generation $g \geq 1$ of the non-root $v$ that topples $n \geq 1$ times by time $t$. Let $u$ be the parent of $v$. Suppose $u$ is the root (i.e., $g = 1$), and suppose (for contradiction) that $u$ topples $m \geq n+2$ times by time $t$. By Theorem 1, we know that $u$ has received either $m$ or $m-1$ grains from each of its children by time $t$, including node $v$. However, by assumption, $v$ topples $n \leq m-2$ times by time $t$, so $u$ receives $\leq m-2$ grains from $v$ by time $t$, a contradiction. Thus, claim (iv) holds for $g = 1$.

Now suppose a node $v$ at generation $g > 1$ topples $n \geq 1$ times by time $t$, and assume that the claim holds at generation $g-1$. Suppose (for contradiction) that $v$'s parent, $u$, toppled $m \geq n+2$ times by time $t$, and let $t' \leq t$ be the moment when $u$ topples for the $m$th time. Before time $t'$, $u$ toppled at most $m-1$ times, so by the inductive hypothesis we know that $u$'s parent has toppled $\leq m$ times. Moreover, by Lemma 2(ii), $u$ receives at least $m$ grains from its parent by time $t'$ because its parent topples an $m$th time. Thus, $u$ receives exactly $m$ grains from its parent by time $t'$, and $u$ topples an $m$th time at time $t'$, so we can apply Lemma 2(iii) to $u$ to conclude that $u$ must have received $m-1$ grains from each of its children (including node $v$) by time $t'$. But $v$ has toppled $n \leq m-2$ times by time $t$, so there is no time $t' \leq t$ at which $u$ receives an $(m-1)$th grain from $v$, a contradiction. Thus, (iv) follows by induction on $g$.

Claim (v) is already shown if $v$ is the root (Theorem 1); here we show the case in which $v$ is non-root by contradiction. If $v$ is not at capacity, then it must receive at least $2$ grains (from any source) before toppling. Consider (for contradiction) the time $t' \leq t$ at which $t$ receives a second grain. Before $t'$, $v$ has toppled $0$ times, so $v$ cannot receive grains from its children [by Lemma 2(i)] and $v$ cannot have received more than one grain from its parent [because $v$'s parent cannot have toppled more than once by Lemma 2(iv)]. Hence, $v$ cannot receive a second grain by time $t'$, a contradiction. So claim (v) holds. □

While Theorem 1 provides strong constraints for the root in any cascade, Theorem 2 provides strong constraints for non-root nodes in cascades forming a finite tree.

**Theorem 2** (Strong constraints on non-root nodes in cascades forming a finite tree). *For a cascade that forms a finite tree $\mathcal{G}^\dagger$, a non-root node $v$ topples $0$ times by time $t > 0$ if and only if $v$ was not initially at capacity or $v$ receives $0$ grains from its parent by time $t$. Moreover, under the same conditions and for any positive integer $n$, a non-root node $v$ topples $n$ times by time $t > 0$ if and only if all of the following conditions hold: (a) $v$ was initially at capacity; (b) $v$ received $n$ or $n+1$ grains from its parent by time $t$; and (c) $v$ received $n$ or $n-1$ grains from each of its children by time $t$ (except not $n$ grains from every child if $v$ received $n+1$ grains from its parent).*

*Proof.* We first prove the first claim concerning $v$ toppling $0$ times. By Lemma 2(v), $v$ topples $0$ times if $v$ is not initially at capacity. Now suppose that the non-root node $v$ receives $0$ grains from its parent by time $t$. By Lemma 2(i) with $n = 0$, $v$ cannot receive a grain from one of its children by time $t$. Thus, $v$ receives a total of $0$ grains by time $t$ and hence topples $0$ times by time $t$.

Inversely, if $v$ is initially at capacity and $v$ receives at least one grain from its parent by time $t$, then $v$ clearly topples at least once by time $t$. This concludes the proof of the claim for $v$ toppling $0$ times.

To prove the second claim, first suppose that conditions (a), (b) and (c) hold. Let $k$ be the degree of $v$. By counting grains, we see that the number of grains initially on $v$ and received by $v$ by time $t$ is in the interval $[kn, k(n+1)-1]$, so $v$ topples $n$ times by time $t$.

To show the converse, suppose one of (a), (b) or (c) does not hold, and suppose (for contradiction) that $v$ topples $n$ times by time $t$. If (a) does not hold (i.e., if $v$ is not initially at capacity), then $v$ topples $0$ times by time $t$ [Lemma 2(v)], a contradiction.

Next consider the two cases in which (b) does not hold. If $v$ received fewer than $n$ grains from its parent by time $t$, then by Lemma 2(ii) $v$ topples fewer than $n$ times by time $t$, a contradiction. If $v$ received $> n+1$ grains from its parent by time $t$ and if $v$'s parent were the root, then by Theorem 1 node $v$ would necessarily topple $\geq n+1$ times by time $t$, a contradiction. Finally, if $v$ received $> n+1$ grains from its parent by time $t$ and if $v$'s parent were not the root, then by part (c) of this theorem applied to the parent of $v$, we know that $v$ topples $\geq n+1$ times by time $t$, a contradiction.

To finish the proof, consider the three ways in which (c) may not hold. If $v$ received fewer than $n-1$ grains from any of its children by time $t$, then $v$ toppled fewer than $n$ times by time $t$ [by Lemma 2(iii)], a contradiction. If $v$ received more than $n$ grains from any of its children by time $t$, then $v$ toppled more than $n$ times by time $t$ [because Lemma 2(ii) implies that such a child must have received more than $n$ grains from its parent, i.e., from $v$], a contradiction. In the last case, $v$ received $n+1$ grains from its parent and $n$ grains from each of its children. Let $k$ be the degree of $v$. By counting grains, we see that the number of grains initially on $v$ and received by $v$ by time $t$ is $k(n+1)$, so $v$ topples $n+1$ times by time $t$, a contradiction. Hence the second claim holds, which completes the proof. □

Together, Theorems 1–2 provide strong constraints for any node in a cascade forming a finite tree, and this at any time during such a cascade. Nonetheless, Corollary 1 turns out to be useful when translating these rules to a branching process. We now prove this corollary.

*Proof of Corollary 1.* We prove claims (i)–(ix) by letting $t$ be some time after the cascade finishes and by applying Theorems 1–2.

We first consider the cases in which $v$ topples $0$ times. Suppose $v$ is the root. By Theorem 1, point (i) is necessary and sufficient for $v$ to topple $0$ times.

Now suppose $v$ is non-root and topples $0$ times. Because $v$ topples zero times, $v$'s parent receive $0$ grains from $v$, so the parent toppled at most $1$ time (by Theorem 1 or Theorem 2 if the parent is the root or not, respectively), and at least one time (otherwise $v$ would not be in $\mathcal{G}'$). Thus, the parent of $v$ has toppled exactly one time. Hence the signature of $v$ is $(1, 0)$. By Theorem 2, $v$ is not initially at capacity, and/or $v$ receives $0$ grains from its parent during the cascade. These conditions leave only 3 possibilities: $v$ is not initially at capacity and receives the grain sent by its parent [point (ii)]; $v$ is not initially at capacity and the grain dissipates [point (iii)]; or $v$ is initially at capacity and the grain dissipates [point (iv)]. There are no other possibilities.

Next consider the cases in which $v$ topples $n \geq 1$ times. If $v$ is the root, then point (v) is equivalent to Theorem 1 and the fact that $v$'s children are the same as $v$'s neighbors.

Now suppose $v$ is not the root. Let $(n', m')$ be the signature of $v$; let $m$ be the amount of sand received by $v$ from its parent; and let $l_c$ be the amount of sand received by $v$ from one of its children $c$ [that child thus has signature $(n, l_c)$]. By Theorem 2, it is necessary and sufficient that: $v$ was initially at capacity, $v$ received $n$ or $n+1$ grains from its parent (so $m \geq n$), and $v$ received from each of its children $n$ or $n-1$ grains (so $n \geq l_c \geq n-1$ for every child $c$ of $v$), except that $v$ cannot receive $n$ grains from all of its children if it received $n+1$ grains from its parent. Moreover, $v$ cannot receive more grains from its parent than the number of times the parent toppled (so $n' \geq m$); $v$'s parent cannot receive more grains from $v$ than the number of times $v$ toppled (so $n \geq m'$); and $v$'s parent must receive at least $n'-1$ grains from $v$ (by Theorem 1 or Theorem 2 if the parent is the root or not, respectively, so $m' \geq n'-1$). Grain exchanges between $v$ and its parent may thus be summarized as $n' \geq m \geq n \geq m' \geq n'-1$, which leaves 4 possibilities (i.e., 4 possible positions of the ">" symbol): $n' > m = n = m' = n'-1$ [the last grain sent by the parent of $v$ toward $v$ dissipated, point (viii)], $n' = m > n = m' = n'-1$ [no dissipation, $v$ cannot receive $n$ grains from

all its children, point (vi)], $n' = m = n > m' = n' - 1$ [the last grain sent by $v$ toward its parent dissipated, point (ix)], and $n' = m = n = m' > n' - 1$ [no dissipation, point (vii)]. There are no other possibilities. □

## SM4. SELF-ORGANIZING MODEL ON 3-REGULAR NETWORKS

The controlled, mechanistic, self-organizing model $\mathcal{M}^\mu_{\text{BTW}}$ of the BTW process $\mathcal{S}^\mu_{\text{BTW}}$ on infinite 3-regular networks takes the form

$$Y^{i'}_i(\mathbf{y}) \equiv \frac{y_{i'}}{y_i} \qquad Z^{i'j'}_{ij}(\mathbf{z}) \equiv \frac{1 + \delta_{i'j'}(z_{i'} - 1)}{1 + \delta_{ij}(z_i - 1)} \qquad L^{i'}_i(\mathbf{z}) = \sum_{j=0}^{2} \phi_{ij} Z^{i'j}_{ij} \qquad \text{(SM6a)}$$

$${}^{i'}_n\!\underset{\sim}{A}(w,x,\mathbf{y},\mathbf{z}) = \begin{cases} \epsilon L^{i'}_2 + \dfrac{\epsilon}{1-\epsilon}{}^{i'}_1\!\underset{\sim}{B} + (1-\epsilon)\sum_{j=0}^{1}\phi_{2j}Y^{j+1}_j Z^{i'(j+1)}_{2j}\left(L^{j+1}_j\right)^2 & \text{for } n = 1 \\ \epsilon_{n-1}^{i'}\!\underset{\sim}{B} + \dfrac{\epsilon}{1-\epsilon}{}^{i'}_n\!\underset{\sim}{B} + (1-\epsilon)^{2n-1}\phi_{22}xw^{n-1}\sum_{j'=1}^{2}\binom{2}{j'-1}Y^{j'}_2 Z^{i'j'}_{22}\left({}^{j'}_{n-1}\!\underset{\sim}{A}\right)^{2-(j'-1)}\left({}^{j'}_{n-1}\!\underset{\sim}{B}\right)^{j'-1} & \text{for } n > 1 \end{cases}$$
(SM6b)

$${}^{i'}_n\!\underset{\sim}{B}(w,x,\mathbf{y},\mathbf{z}) = (1-\epsilon)^{2n}\phi_{22}xw^n \sum_{j'=0}^{2}\binom{2}{j'}Y^{j'}_2 Z^{i'j'}_{22}\left({}^{j'}_n\!\underset{\sim}{A}\right)^{2-j'}\left({}^{j'}_n\!\underset{\sim}{B}\right)^{j'} \qquad \text{(SM6c)}$$

$$H(w,x,\mathbf{y},\mathbf{z}) = \frac{1-\mu}{1-\psi_2}\sum_{i=0}^{1}\psi_i Y^{i+1}_i\left(L^{i+1}_i\right)^3 + \mu x \sum_{n=1}^{\infty} w^n \sum_{i'=0}^{2}\binom{3}{i'}Y^{i'}_2 \left({}^{i'}_n\!\underset{\sim}{A}\right)^{3-i'}\left({}^{i'}_n\!\underset{\sim}{B}\right)^{i'}. \qquad \text{(SM6d)}$$

This section shows how to obtain this closed system of equations using Corollary 1.

We see from Corollary 1 that the only possible signatures for non-root patterns have the form $(n, n-1)$ or $(n, n)$, where $n \geq 1$ is the number of times that the parent topples. From the definitions of ${}^{i'}_n\!\underset{\sim}{A}$ and ${}^{i'}_n\!\underset{\sim}{B}$ in the main article, we see that ${}^{i'}_n\!\underset{\sim}{A}$ [resp. ${}^{i'}_n\!\underset{\sim}{B}$] tracks the contribution of a non-leaf motif with signature $(n, n-1)$ [resp. $(n, n)$] and of all its descendants. Similarly, $H(w, x, \mathbf{y}, \mathbf{z})$ tracks the contribution of the root pattern and of all its descendants.

For convenience, Eq. (SM6a) defines functions simplifying how the sought PGF depends on the generators $y_i$ and $z_i$. A factor of $Y^{i'}_i(\mathbf{y})$ should be included whenever an $i$-sand nodes becomes $i'$-sand: each pattern (root or non-root) is "responsible" for tracking the factor $Y^{i'}_i(\mathbf{y})$ resulting from a change in the amount of sand on the node associated to this pattern. Similarly, a factor of $Z^{i'j'}_{ij}(\mathbf{z})$ should be included whenever an $ij$-sand link becomes $i'j'$-sand: non-root patterns (whether they topple or not) are responsible for tracking the factor $Z^{i'j'}_{ij}(\mathbf{z})$ due to the link joining them to their parent. This last scheme "forgets" to track changes in *leaf links*, which join a node with an associated pattern to a node not associated to any pattern (hence forming the "leaves" of the branching process). A factor of $L^{i'}_i(\mathbf{z})$ should be included for each node not associated to a pattern that is neighbor to an $i$-sand node that becomes $i'$-sand: all patterns (root or non-root) that do not topple are responsible for tracking such a factor $L^{i'}_i(\mathbf{z})$ for each of their neighbors not associated to a pattern. Finally, each pattern (root or non-root) that topples $n \geq 1$ times is responsible for tracking the factor $xw^n$, because it contributes 1 to the cascade area and $n$ to the cascade size.

We first focus on ${}^{i'}_n\!\underset{\sim}{B}(w, x, \mathbf{y}, \mathbf{z})$ [Eq. (SM6c)], which corresponds to the pattern (vii) in Fig. SM3 and in Corollary 1: a non-root node receives from and returns to its parent $n$ grains, and the parent ends up as $i'$-sand after the cascade. The factor $(1-\epsilon)^{2n}$ accounts for the probability that all the grains traveling from and to the parent do not dissipate, while $\phi_{22}$ accounts for the probability that this node is 2-sand at the beginning of the cascade (the parent was initially 2-sand because it topples at least once). The two children patterns of this pattern (vii) may have any combination of signature $(n, n-1)$ or $(n, n)$, so suppose $j' \in \{0, 1, 2\}$ of them have signature $(n, n)$ and $2 - j'$ thus have signature $(n, n-1)$. Summing the 2 grains initially on this node, the $n$ grains received from its parent, and the $nj' + (n-1)(2-j')$ received from its children, we obtain $3n + j'$, enough to topple $n$ times and ending up $j'$-sand. Hence, this pattern is responsible for tracking $xw^n Y^{j'}_2 Z^{i'j'}_{22}$, and it defers to its children the task of tracking $\left({}^{j'}_n\!\underset{\sim}{A}\right)^{2-j'}\left({}^{j'}_n\!\underset{\sim}{B}\right)^{j'}$. A combinatorial factor $\binom{2}{j'}$ accounts from the number of different ways to choose the signature of the children.

We now focus on ${}^{i'}_1\!\underset{\sim}{A}(w, x, \mathbf{y}, \mathbf{z})$ [Eq. (SM6b) with $n = 1$], which corresponds to a non-root node receiving 1 grain from its parent, returning no grains, and whose parent end up being $i'$-sand after the cascade. The first term, $\epsilon L^{i'}_2$, accounts for the probability $\epsilon$ that the received grain dissipates before reaching the child [pattern (iii) or (iv)], which has the same contribution as that of a leaf whose parent transitioned from being 2-sand to $i'$-sand. The second term considers the possibility that the child

topples, but the grain sent to its parent dissipates before reaching it [pattern (ix)]: expanding that term gives

$$\frac{\epsilon}{1-\epsilon}{}_1^{i'}\!B = (1-\epsilon)\epsilon\phi_{22}xw\sum_{j'=0}^{2}\binom{2}{j'}Y_2^{j'}Z_{22}^{i'j'}\big({}_1^{j'}\!A\big)^{2-j'}\big({}_1^{j'}\!B\big)^{j'}, \qquad (SM7)$$

where the factor $(1-\epsilon)\epsilon$ accounts for the probability that this node receives the grain sent by its parent, but not the converse; and the rest of the expression is obtained in the same way as ${}_1^{i'}\!B(w,x,\mathbf{y},\mathbf{z})$. Finally, the third term of ${}_1^{i'}\!A(w,x,\mathbf{y},\mathbf{z})$ corresponds to the pattern (ii): this node receives the grain sent by its parent [factor $(1-\epsilon)$], but it does not topple because it is initially $j$-sand with $j \in \{0,1\}$ (i.e., not at capacity). This event, which has probability $\phi_{2j}$ to occur, mandates the contribution $xw^n Y_j^{j+1} Z_{2j}^{i'(j+1)}\big(L_j^{j+1}\big)^2$ for which this pattern is responsible.

The case ${}_n^{i'}\!A(w,x,\mathbf{y},\mathbf{z})$ with $n > 1$ [Eq. (SM6b)] corresponds to a non-root node that receives $n$ grains from its parent, that returns $n-1$ grains, and whose parent end up being $i'$-sand after the cascade. Here the first term corresponds to the case in which the last grain sent by the parent dissipates [pattern (viii)], which amounts to ${}_{n-1}^{i'}\!B$ with a factor $\epsilon$ for the extra dissipation. Similarly, the second term corresponds to the case in which the last grain sent by this node to its parent dissipates [pattern (ix)], which amounts to ${}_n^{i'}\!B$ with a factor $\epsilon/(1-\epsilon)$ rectifying the probability for one extra sand dissipation and one fewer successful sand transfer. The last term corresponds to the pattern (vi): none of the grains exchanged between this node and its parent dissipate [factor $(1-\epsilon)^{2n-1}$]; this node is initially at capacity (factor $\phi_{22}$); and this node topples $n-1$ times (factor $xw^{n-1}$). The two children of this pattern (vi) may have signature $(n-1, n-2)$ or $(n-1, n-1)$, but not both of them may have signature $(n-1, n-1)$, so suppose $j'-1 \in \{0,1\}$ of them have signature $(n-1, n-1)$ and $2-(j'-1)$ have signature $(n-1, n-2)$. Summing the 2 grains initially on this node, the $n$ grains received from its parent, and the $(n-1)(j'-1) + (n-2)[2-(j'-1)]$ received from its children, we obtain $3(n-1) + j'$ grains, enough for this node to topple $n-1$ times and end up $j'$-sand. Hence, this pattern is responsible for tracking $xw^{n-1}Y_2^{j'}Z_{22}^{i'j'}$, and it defers to its children the task of tracking $\big({}_{n-1}^{j'}\!A\big)^{2-(j'-1)}\big({}_{n-1}^{j'}\!B\big)^{j'-1}$. A combinatorial factor $\binom{2}{j'-1}$ accounts from the number of different ways to choose the signature of the children.

Finally, $H(w,x,\mathbf{y},\mathbf{z})$ [Eq. (SM6d)] tracks the contribution of root patterns and of their descendants. The first term corresponds to the case in which the root is $i$-sand with $i \in \{0,1\}$ (total probability $1-\mu$; each outcome $i \in \{0,1\}$ has weight $\psi_i$; and $\psi_0 + \psi_1 = 1 - \psi_2$), so the root does not topple [pattern (i)]: the pattern is then responsible for a factor $Y_i^{i+1}\big(L_i^{i+1}\big)^3$. The second term accounts for the root being initially at capacity (probability $\mu$) and toppling $n$ times: the three children of this pattern (v) may have signature $(n, n-1)$ or $(n, n)$, but not all of them may have signature $(n,n)$, so suppose $i' \in \{0,1,2\}$ of them have signature $(n,n)$ and $3-i'$ have signature $(n, n-1)$. Summing the 2 grains initially on this node, the 1 grain dropped on this root to start the cascade, and the $ni' + (n-1)(3-i')$ received from its children, we obtain $3n + i'$ grains, enough for this node to topple $n$ times and end up $i'$-sand. Hence, this pattern is responsible for tracking $xw^n Y_2^{i'}$, and defers to its children the task of tracking $\big({}_n^{i'}\!A\big)^{3-i'}\big({}_n^{i'}\!B\big)^{i'}$. A combinatorial factor $\binom{3}{i'}$ accounts for the number of different ways to choose the signatures of the children of the root.

Now we check that setting

$$w = y_0 = y_1 = y_2 = z_0 = z_1 = z_2 = 1 \qquad (SM8)$$

in Eq. (SM6) recovers $\mathcal{E}^\mu_{\text{BTW}}$ [i.e., Eq. (1) of the main article]

$$F(x) = 1 - (1-\epsilon)\phi_{22} + (1-\epsilon)\phi_{22}x\big[F(x)\big]^2 \qquad (SM9a)$$

$$G(x) = 1 - \mu + \mu x\big[F(x)\big]^3, \qquad (SM9b)$$

a form with which the reader may be more familiar. The substitution (SM8) gives $Y_i^{i'}(\mathbf{1}) = Z_{ij}^{i'j'}(\mathbf{1}) = L_i^{i'}(\mathbf{1}) = 1$ and

$${}_1^*\!A(x) \equiv {}_1^{i'}\!A(1,x,\mathbf{1},\mathbf{1}) = \epsilon + \frac{\epsilon}{1-\epsilon}{}_1^*\!B + (1-\epsilon)(1-\phi_{22}) \qquad (SM10a)$$

$$\begin{aligned}{}_{n>1}^*\!A(x) \equiv {}_{n>1}^{i'}\!A(1,x,\mathbf{1},\mathbf{1}) &= \epsilon\,{}_{n-1}^*\!B + \frac{\epsilon}{1-\epsilon}{}_n^*\!B + (1-\epsilon)^{2n-1}\phi_{22}x\sum_{j'=1}^{2}\binom{2}{j'-1}\big({}_{n-1}^*\!A\big)^{2-(j'-1)}\big({}_{n-1}^*\!B\big)^{j'-1} \\ &= \epsilon\,{}_{n-1}^*\!B + \frac{\epsilon}{1-\epsilon}{}_n^*\!B + (1-\epsilon)^{2n-1}\phi_{22}x\Big[\big({}_{n-1}^*\!A + {}_{n-1}^*\!B\big)^2 - \big({}_{n-1}^*\!B\big)^2\Big]\end{aligned} \qquad (SM10b)$$

$${}_n^*\!B(x) \equiv {}_n^{i'}\!B(1,x,\mathbf{1},\mathbf{1}) = (1-\epsilon)^{2n}\phi_{22}x\sum_{j'=0}^{2}\binom{2}{j'}\big({}_n^*\!A\big)^{2-j'}\big({}_n^*\!B\big)^{j'} = (1-\epsilon)^{2n}\phi_{22}x\big({}_n^*\!A + {}_n^*\!B\big)^2 \qquad (SM10c)$$

$$H(1,x,\mathbf{1},\mathbf{1}) = 1 - \mu + \mu x\sum_{n=1}^{\infty}\sum_{i'=0}^{2}\binom{3}{i'}\big({}_n^*\!A\big)^{3-i'}\big({}_n^*\!B\big)^{i'} = 1 - \mu + \mu x\sum_{n=1}^{\infty}\Big[\big({}_n^*\!A + {}_n^*\!B\big)^3 - \big({}_n^*\!B\big)^3\Big], \qquad (SM10d)$$

where we used the fact that ${}_n^{i'}\!A(1,x,\mathbf{1},\mathbf{1})$ and ${}_n^{i'}\!B(1,x,\mathbf{1},\mathbf{1})$ are not affected by changes in $i'$ (so we replaced $i'$ by $*$). For $n > 1$, we may verify that the ansatz ${}_{n-1}^{*}B(x) = {}_n^{*}A(x) + {}_n^{*}B(x)$ satisfies the equations for ${}_n^{*}A(x)$ and ${}_n^{*}B(x)$ [Eqs. (SM10a)–(SM10c)], which provides $F(x) = {}_1^{*}A(x) + {}_1^{*}B(x)$. Furthermore, the same ansatz provides $G(x) = \tilde{H}(1,x,\mathbf{1},\mathbf{1})$ through a telescoping series. Thus, substitution (SM8) reduces the mechanistic model Eq. (SM10) to the single-type branching process for cascade area Eq. (SM9).

## SM5. THE DERIVATIVES USED IN THE BOOTSTRAPPING

This section explicitly shows how one may evaluate the derivatives required to enforce the equilibrium of the mechanistic model $\mathcal{M}_{\mathrm{BTW}}^{\mu}$. To this end, we consider the following equations and definitions

$$1 = \left. Y_i^{i'} \right|_1 \qquad 1 = \left. Z_{ij}^{i'j'} \right|_1 \qquad 1 = \left. L_i^{i'} \right|_1 \qquad {}_n\mathcal{A} \equiv \left. {}_n^{i'}\!A \right|_1 \qquad {}_n\mathcal{B} \equiv \left. {}_n^{i'}\!B \right|_1 \qquad \mathcal{H} \equiv \left. H \right|_1$$

$$\gamma_{i(k)}^{i'} \equiv \left. \frac{\partial Y_i^{i'}}{\partial y_k} \right|_1 \qquad 0 = \left. \frac{\partial Z_{ij}^{i'j'}}{\partial y_k} \right|_1 \qquad 0 = \left. \frac{\partial L_i^{i'}}{\partial y_k} \right|_1 \qquad {}_n a_{(k)} \equiv \left. \frac{\partial {}_n^{i'}\!A}{\partial y_k} \right|_1 \qquad {}_n b_{(k)} \equiv \left. \frac{\partial {}_n^{i'}\!B}{\partial y_k} \right|_1 \qquad h_{(k)} \equiv \left. \frac{\partial H}{\partial y_k} \right|_1$$

$$0 = \left. \frac{\partial Y_i^{i'}}{\partial z_k} \right|_1 \qquad \zeta_{ij(k)}^{i'j'} \equiv \left. \frac{\partial Z_{ij}^{i'j'}}{\partial z_k} \right|_1 \qquad \lambda_{i(k)}^{i'} \equiv \left. \frac{\partial L_i^{i'}}{\partial z_k} \right|_1 \qquad {}_n^{i'}\!\alpha_{(k)} \equiv \left. \frac{\partial {}_n^{i'}\!A}{\partial z_k} \right|_1 \qquad {}_n^{i'}\!\beta_{(k)} \equiv \left. \frac{\partial {}_n^{i'}\!B}{\partial z_k} \right|_1 \qquad \eta_{(k)} \equiv \left. \frac{\partial H}{\partial z_k} \right|_1$$

where the notation $\bullet|_1$ means $\bullet|_{w=1,x=1,\mathbf{y}=\mathbf{1},\mathbf{z}=\mathbf{1}}$. Whenever an index appear on the right hand side and not on the left hand side, it means that the value of the right hand side is the same irrespective of the value of the index. The first row evaluates important objects when all the generators take unit value, while the second and third rows consider derivatives with respect to $y_k$ and $z_k$ at the same point, respectively. Some of these derivatives are trivial (i.e., equal to 1 or 0); we simply give their value on the left hand side. Nontrivial objects are named. We seek the values of all these quantities, including the $h_{(k)}$ and $\eta_{(k)}$ required for the constraints in the main article.

The three nontrivial objects evaluated at one without differentiation are

$$\begin{aligned}
{}_1\mathcal{A} &= \epsilon + \frac{\epsilon\,{}_1\mathcal{B}}{1-\epsilon} + (1-\epsilon)(\phi_{20} + \phi_{21}) = 1 - (1-\epsilon)\phi_{22} + \frac{\epsilon\,{}_1\mathcal{B}}{1-\epsilon} \\
{}_{n\neq 1}\mathcal{A} &= \epsilon\,{}_{n-1}\mathcal{B} + \frac{\epsilon\,{}_n\mathcal{B}}{1-\epsilon} + (1-\epsilon)^{2n-1}\phi_{22}\left[({}_{n-1}\mathcal{A})^2 + 2\,{}_{n-1}\mathcal{A}\,{}_{n-1}\mathcal{B}\right] \\
{}_n\mathcal{B} &= (1-\epsilon)^{2n}\phi_{22}\left[({}_n\mathcal{A})^2 + 2\,{}_n\mathcal{A}\,{}_n\mathcal{B} + ({}_n\mathcal{B})^2\right] = (1-\epsilon)^{2n}\phi_{22}\left[{}_n\mathcal{A} + {}_n\mathcal{B}\right]^2 \\
\mathcal{H} &= \psi_0 + \psi_1 + \psi_2 \sum_{n=1}^{\infty}\left[({}_n\mathcal{A})^3 + 3({}_n\mathcal{A})^2\,{}_n\mathcal{B} + 3\,{}_n\mathcal{A}({}_n\mathcal{B})^2\right] = 1 - \psi_2 + \psi_2 \sum_{n=1}^{\infty}\left[({}_n\mathcal{A} + {}_n\mathcal{B})^3 - ({}_n\mathcal{B})^3\right].
\end{aligned}$$

If the cascades are of finite size (i.e., the branching factor $R_0$, the mean number of topplings caused by a toppling, is $\leq 1$, which includes the case of interest $R_0 = 1$), then the solution to this system is

$${}_n\mathcal{A} = {}_{n-1}\mathcal{B} - {}_n\mathcal{B} \qquad\qquad {}_n\mathcal{B} = (1-\epsilon)^{2^{n+2}-2n-4}(\phi_{22})^{2^n-1} \qquad\qquad \mathcal{H} = 1.$$

Note that this solution is also valid for ${}_1\mathcal{A}$ if we allow ${}_0\mathcal{B} = 1$.

Differentiation with respect to $y_k$ followed by evaluation at one gives rise to four nontrivial objects

$$\begin{aligned}
\gamma_{i(k)}^{i'} &= \delta_{i'k} - \delta_{ik} \\
{}_1 a_{(k)} &= \frac{\epsilon\,{}_1 b_{(k)}}{1-\epsilon} + (1-\epsilon)\left(\phi_{20}\gamma_{0(k)}^1 + \phi_{21}\gamma_{1(k)}^2\right) \\
{}_{n\neq 1} a_{(k)} &= \epsilon\,{}_{n-1} b_{(k)} + \frac{\epsilon\,{}_n b_{(k)}}{1-\epsilon} + (1-\epsilon)^{2n-1}\phi_{22}\left[{}_{n-1}\mathcal{A}\left({}_{n-1}\mathcal{A}\gamma_{2(k)}^1 + 2\,{}_{n-1}a_{(k)}\right) + 2\left({}_{n-1}\mathcal{B}\,{}_{n-1}a_{(k)} + {}_{n-1}\mathcal{A}\,{}_{n-1}b_{(k)}\right)\right] \\
{}_n b_{(k)} &= (1-\epsilon)^{2n}\phi_{22}\Big[\gamma_{2(k)}^0({}_n\mathcal{A})^2 + 2\,{}_n a_{(k)}\,{}_n\mathcal{A} + 2\gamma_{2(k)}^1\,{}_n\mathcal{A}\,{}_n\mathcal{B} + 2\,{}_n a_{(k)}\,{}_n\mathcal{B} + 2\,{}_n\mathcal{A}\,{}_n b_{(k)} + 2\,{}_n b_{(k)}\,{}_n\mathcal{B}\Big] \\
h_{(k)} &= \psi_0\gamma_{0(k)}^1 + \psi_1\gamma_{1(k)}^2 + \psi_2 \sum_{n=1}^{\infty}\Big[\gamma_{2(k)}^0({}_n\mathcal{A})^3 + 3\,{}_n a_{(k)}({}_n\mathcal{A})^2 + 3\gamma_{2(k)}^1({}_n\mathcal{A})^2\,{}_n\mathcal{B} + 6\,{}_n a_{(k)}\,{}_n\mathcal{A}\,{}_n\mathcal{B} \\
&\qquad\qquad\qquad\qquad\qquad + 3({}_n\mathcal{A})^2\,{}_n b_{(k)} + 3\,{}_n a_{(k)}({}_n\mathcal{B})^2 + 6\,{}_n\mathcal{A}\,{}_n b_{(k)}\,{}_n\mathcal{B}\Big].
\end{aligned}$$

Suppose we fix $k$. We notice that $_1a_{(k)}$ and $_1b_{(k)}$ only depend on $\gamma^{i'}_{i(k)}$ (which are easily evaluated); on $_n\mathcal{A}$ and $_n\mathcal{B}$ (for which we already know the value at this point); and on $_1a_{(k)}$ and $_1b_{(k)}$ themselves. We thus solve this linear system of two equations and two unknowns. For a fixed $n$, assuming that we know $_{n-1}a_{(k)}$ and $_{n-1}b_{(k)}$, we see that $_na_{(k)}$ and $_nb_{(k)}$ only depend on known quantities and on themselves. We thus iteratively solve linear systems and obtain $_na_{(k)}$ and $_nb_{(k)}$ for values of $n$ as large as we need. This iterative procedure allows the estimation of $h_{(k)}$ by choosing some finite truncation value $n_{\max}$ for the upper bound of the summation. We then repeat the procedure for a different value of $k$.

Differentiation with respect to $z_k$ followed by evaluation at one gives the five nontrivial objects

$$\zeta^{i'j'}_{ij(k)} = \delta_{i'k}\delta_{i'j'} - \delta_{ik}\delta_{ij}$$

$$\lambda^{i'}_{i(k)} = \sum_{j=0}^{2} \phi_{ij}(\delta_{i'k}\delta_{i'j} - \delta_{ik}\delta_{ij}) = \delta_{i'k}\phi_{ii'} - \delta_{ik}\phi_{ii}$$

$$_1\alpha^{i'}_{(k)} = \epsilon\lambda^{i'}_{2(k)} + \frac{\epsilon\, _1\beta^{i'}_{(k)}}{1-\epsilon} + (1-\epsilon)\left[\phi_{20}\left(\zeta^{i'1}_{20(k)} + 2\lambda^1_{0(k)}\right) + \phi_{21}\left(\zeta^{i'2}_{21(k)} + 2\lambda^2_{1(k)}\right)\right]$$

$$_{n\neq 1}\alpha^{i'}_{(k)} = \epsilon\, _{n-1}\beta^{i'}_{(k)} + \frac{\epsilon\, _n\beta^{i'}_{(k)}}{1-\epsilon} + (1-\epsilon)^{2n-1}\phi_{22}\Big[\zeta^{i'1}_{22(k)}({}_{n-1}\mathcal{A})^2 + 2\, _{n-1}\alpha^1_{(k)}{}_{n-1}\mathcal{A} + 2\zeta^{i'2}_{22(k)}{}_{n-1}\mathcal{A}{}_{n-1}\mathcal{B}$$
$$+ 2\, _{n-1}\alpha^2_{(k)}{}_{n-1}\mathcal{B} + 2\, _{n-1}\mathcal{A}{}_{n-1}\beta^2_{(k)}\Big]$$

$$_n\beta^{i'}_{(k)} = (1-\epsilon)^{2n}\phi_{22}\Big[\zeta^{i'0}_{22(k)}({}_n\mathcal{A})^2 + 2\, _n\alpha^0_{(k)}{}_n\mathcal{A} + 2\zeta^{i'1}_{22(k)}{}_n\mathcal{A}{}_n\mathcal{B} + 2\, _n\alpha^1_{(k)}{}_n\mathcal{B} + 2\, _n\mathcal{A}{}_n\beta^1_{(k)} + \zeta^{i'2}_{22(k)}({}_n\mathcal{B})^2 + 2\, _n\beta^2_{(k)}{}_n\mathcal{B}\Big]$$

$$\eta_{(k)} = 3\psi_0\lambda^1_{0(k)} + 3\psi_1\lambda^2_{1(k)} + \psi_2\sum_{n=1}^{\infty}\Big[3\, _n\alpha^0_{(k)}({}_n\mathcal{A})^2 + 6\, _n\alpha^1_{(k)}{}_n\mathcal{A}{}_n\mathcal{B} + 3({}_n\mathcal{A})^2\, _n\beta^1_{(k)} + 3\, _n\alpha^2_{(k)}({}_n\mathcal{B})^2 + 6\, _n\mathcal{A}{}_n\beta^2_{(k)}{}_n\mathcal{B}\Big].$$

As before, we can fix $k$, start with $n = 1$, and iteratively solve for $_n\alpha^{i'}_{(k)}$ and $_n\beta^{i'}_{(k)}$. This time, due to the index $i'$, the linear system contains six equations and six unknowns. Nonetheless, we can still solve for as large a value of $n$ as desired to estimate $\eta_{(k)}$.

## SM6. VALUES OF $\psi_i$ AND $\phi_{ij}$

Table SM1 provides estimations of $\psi_i$ and $\phi_{ij}$ from Monte Carlo simulations of the controlled BTW sandpile process on a 3-regular network, $\mathcal{S}^\mu_{\text{BTW}}$, and from the controlled, self-organizing, mechanistic, analytical model, $\mathcal{M}^\mu_{\text{BTW}}$. The entries for the latter correspond to the numerical solution of the analytical system of constraints

$$h_{(k)} = 0 \;\forall k \in \{0,1,2\} \tag{SM11a}$$

$$\eta_{(k)} = 0 \;\forall k \in \{0,1,2\} \tag{SM11b}$$

$$\sum_{i=0}^{2} \psi_i = 1 \tag{SM11c}$$

$$\sum_{j=0}^{2} \phi_{ij} = 1 \;\forall i \in \{0,1,2\} \tag{SM11d}$$

$$\psi_i\phi_{ij} = \psi_j\phi_{ji} \;\forall i,j \in \{0,1,2\}. \tag{SM11e}$$

Equation (SM11a) ensures that the number of $i$-sand nodes is at a dynamical equilibrium and provides 2 independent constraints (because $h_{(0)} + h_{(1)} + h_{(2)} = 0$ by construction); Eq. (SM11b) ensures that the number of $ij$-sand links is at a dynamical equilibrium and provides 3 independent constraints; Eq. (SM11c) ensures that the $\psi_i$ are normalized and provides 1 independent constraint; Eq. (SM11d) ensures that the $\phi_{ij}$ are normalized and provides 3 independent constraints; and Eq. (SM11e) ensures that the $\psi_i$ and $\phi_{ij}$ are consistent among themselves (Bayes' rule) and provides 3 independent constraints. Hence, there are 12 independent constraints on the 12 unknowns $\psi_i$ and $\phi_{ij}$. Because Eqs. (SM11a)–(SM11b) are non-linear, *a priori* there is no guarantee that a valid solution exists, nor that there is a single solution. However, starting from educated guesses, sensible numerical solutions are found.

Qualitatively, $\mathcal{M}^\mu_{\text{BTW}}$ shows great prediction of the behavior of $\mathcal{S}^\mu_{\text{BTW}}$ over a wide range of parameters $\epsilon$ and $\mu$. Quantitatively, the values provided by $\mathcal{M}^\mu_{\text{BTW}}$ are in acceptable agreement with those seen in $\mathcal{S}^\mu_{\text{BTW}}$ for large $N$, although the results for $\mathcal{S}^\mu_{\text{BTW}}$ may converge to values different from those of $\mathcal{M}^\mu_{\text{BTW}}$ in the limit $N \to \infty$. This deviation is likely due to the presence of correlations in $\mathcal{S}^\mu_{\text{BTW}}$ that are neglected in $\mathcal{M}^\mu_{\text{BTW}}$. More precisely, $\mathcal{M}^\mu_{\text{BTW}}$ only consider dyadic correlations (through $\phi_{ij}$),

TABLE SM1. Approximate values of $\psi_i$ and $\phi_{ij}$. Each row corresponds to a specific choice of $\epsilon$ and $\mu$. When "$\to 0$" appears in the column labeled $\mu$, no control is applied. When "$\psi_2$" appears in the column labeled $\epsilon$, the limit $\epsilon \to 0$ is taken (which we obtain by linear extrapolation). Values of $\psi_i$ (given in vector form) and of $\phi_{ij}$ (given in matrix form) are obtained through Monte Carlo simulations of $\mathcal{S}_{\mathrm{BTW}}^\mu$, for different network sizes $N$ (using the algorithm described in Sec. SM1 with Convergence_iterations = $10N$, Inner_statistics_iterations = $10^4$ and Outer_statistics_iterations = $10^{10}/N$, so $10^{10}$ data points were obtained in each case) and by the numerical solution of the mechanistic, self-organizing model $\mathcal{M}_{\mathrm{BTW}}^\mu$ (which assumes $N \to \infty$). The standard error is 0.00002 or better. The qualitative behavior of the values obtained with the mechanistic, self-organizing model $\mathcal{M}_{\mathrm{BTW}}^\mu$ is in great agreement with the Monte Carlo simulations $\mathcal{S}_{\mathrm{BTW}}^\mu$, and this agreement holds over a vast range of parameters $\epsilon$ and $\mu$. Although the quantitative agreement is acceptable, the results for $\mathcal{S}_{\mathrm{BTW}}^\mu$ may converge in the limit $N \to \infty$ to values different from those predicted by $\mathcal{M}_{\mathrm{BTW}}^\mu$. Extra correlations present in $\mathcal{S}_{\mathrm{BTW}}^\mu$ and neglected in $\mathcal{M}_{\mathrm{BTW}}^\mu$ may explain this deviation.

| $\epsilon$ | $\mu$ | $\begin{pmatrix}\psi_0\\\psi_1\\\psi_2\end{pmatrix}$ Monte Carlo simulations of $\mathcal{S}_{\mathrm{BTW}}^\mu$ | | | $\mathcal{M}_{\mathrm{BTW}}^\mu$ | $\begin{pmatrix}\phi_{00}&\phi_{01}&\phi_{02}\\\phi_{10}&\phi_{11}&\phi_{12}\\\phi_{20}&\phi_{21}&\phi_{22}\end{pmatrix}$ Monte Carlo simulations of $\mathcal{S}_{\mathrm{BTW}}^\mu$ | | | $\mathcal{M}_{\mathrm{BTW}}^\mu$ |
|---|---|---|---|---|---|---|---|---|---|
| | | $N=10^5$ | $N=10^6$ | $N=10^7$ | $N\to\infty$ | $N=10^5$ | $N=10^6$ | $N=10^7$ | $N\to\infty$ |
| 0.05 | $\psi_2$ | $\begin{pmatrix}0.11351\\0.35651\\0.52998\end{pmatrix}$ | $\begin{pmatrix}0.11350\\0.35651\\0.52999\end{pmatrix}$ | $\begin{pmatrix}0.11352\\0.35650\\0.52998\end{pmatrix}$ | $\begin{pmatrix}0.12117\\0.36700\\0.51183\end{pmatrix}$ | $\begin{pmatrix}0.02626&0.30673&0.66701\\0.09767&0.34746&0.55487\\0.14286&0.37324&0.48389\end{pmatrix}$ | $\begin{pmatrix}0.02627&0.30668&0.66705\\0.09766&0.34745&0.44389\\0.14285&0.37324&0.48391\end{pmatrix}$ | $\begin{pmatrix}0.02625&0.30671&0.66704\\0.09767&0.34753&0.55480\\0.14287&0.37321&0.48392\end{pmatrix}$ | $\begin{pmatrix}0.02615&0.32497&0.64888\\0.10730&0.36413&0.52858\\0.15362&0.37901&0.46737\end{pmatrix}$ |
| 0.01 | $\psi_2$ | $\begin{pmatrix}0.08984\\0.33939\\0.57077\end{pmatrix}$ | $\begin{pmatrix}0.08985\\0.33947\\0.57069\end{pmatrix}$ | $\begin{pmatrix}0.08983\\0.33950\\0.57067\end{pmatrix}$ | $\begin{pmatrix}0.10063\\0.35725\\0.54212\end{pmatrix}$ | $\begin{pmatrix}0.00507&0.26409&0.73085\\0.06991&0.32165&0.60844\\0.11505&0.36179&0.52316\end{pmatrix}$ | $\begin{pmatrix}0.00507&0.26409&0.73084\\0.06990&0.32172&0.60838\\0.11506&0.36190&0.52305\end{pmatrix}$ | $\begin{pmatrix}0.00507&0.26406&0.73086\\0.06989&0.32175&0.60836\\0.11505&0.36191&0.52304\end{pmatrix}$ | $\begin{pmatrix}0.00506&0.28864&0.70630\\0.08130&0.34794&0.57076\\0.13110&0.37612&0.49278\end{pmatrix}$ |
| 0.001 | $\psi_2$ | $\begin{pmatrix}0.08400\\0.33337\\0.58263\end{pmatrix}$ | $\begin{pmatrix}0.08400\\0.33392\\0.58209\end{pmatrix}$ | $\begin{pmatrix}0.08397\\0.33403\\0.58199\end{pmatrix}$ | $\begin{pmatrix}0.09574\\0.35439\\0.54987\end{pmatrix}$ | $\begin{pmatrix}0.00050&0.25148&0.74802\\0.06336&0.31295&0.62368\\0.10784&0.35687&0.53529\end{pmatrix}$ | $\begin{pmatrix}0.00050&0.25148&0.74802\\0.06326&0.31342&0.62331\\0.10794&0.35758&0.53448\end{pmatrix}$ | $\begin{pmatrix}0.00050&0.25148&0.74802\\0.06321&0.31349&0.62330\\0.10795&0.35773&0.53432\end{pmatrix}$ | $\begin{pmatrix}0.00050&0.27802&0.72148\\0.07511&0.34286&0.58203\\0.12562&0.37512&0.49926\end{pmatrix}$ |
| $\to 0$ | $\psi_2$ | — | — | — | $\begin{pmatrix}0.09519\\0.35406\\0.55075\end{pmatrix}$ | — | — | — | $\begin{pmatrix}0.00000&0.27676&0.72324\\0.07441&0.34226&0.58334\\0.12500&0.37500&0.50000\end{pmatrix}$ |
| 0.05 | 0.05 | $\begin{pmatrix}0.09699\\0.34546\\0.55754\end{pmatrix}$ | $\begin{pmatrix}0.09709\\0.34558\\0.55733\end{pmatrix}$ | $\begin{pmatrix}0.09709\\0.34558\\0.55733\end{pmatrix}$ | $\begin{pmatrix}0.10526\\0.34160\\0.55314\end{pmatrix}$ | $\begin{pmatrix}0.02378&0.28439&0.69183\\0.07985&0.33818&0.58197\\0.12036&0.36060&0.51904\end{pmatrix}$ | $\begin{pmatrix}0.02380&0.28463&0.69158\\0.07997&0.33838&0.58165\\0.12047&0.36065&0.51887\end{pmatrix}$ | $\begin{pmatrix}0.02382&0.28473&0.69144\\0.07999&0.33845&0.58156\\0.12048&0.36069&0.51884\end{pmatrix}$ | $\begin{pmatrix}0.02393&0.29686&0.67921\\0.09147&0.34124&0.56728\\0.12925&0.35034&0.52041\end{pmatrix}$ |
| 0.05 | 0.99 | $\begin{pmatrix}0.12674\\0.37005\\0.50322\end{pmatrix}$ | $\begin{pmatrix}0.12674\\0.37005\\0.50321\end{pmatrix}$ | $\begin{pmatrix}0.12674\\0.37010\\0.50316\end{pmatrix}$ | $\begin{pmatrix}0.13615\\0.39157\\0.47228\end{pmatrix}$ | $\begin{pmatrix}0.02829&0.32659&0.64513\\0.11185&0.35810&0.53005\\0.16249&0.38978&0.44774\end{pmatrix}$ | $\begin{pmatrix}0.02830&0.32659&0.64511\\0.11185&0.35810&0.53006\\0.16249&0.38978&0.44773\end{pmatrix}$ | $\begin{pmatrix}0.02833&0.32664&0.64503\\0.11186&0.35809&0.53005\\0.16247&0.38987&0.44766\end{pmatrix}$ | $\begin{pmatrix}0.02818&0.35328&0.61854\\0.12284&0.38777&0.48939\\0.17831&0.40576&0.41593\end{pmatrix}$ |

while correlations between the amount of sand on nodes arbitrarily far from one another could appear in the network of $\mathcal{S}_{\text{BTW}}^\mu$ (especially in the limit $\epsilon \to 0$). For example, the model $\mathcal{M}_{\text{BTW}}^\mu$ assumes that the probability that an $i$-sand node chosen uniformly at random has $m$ many 0-sand neighbors, $n$ many 1-sand neighbors and $(3-m-n)$ many 2-sand neighbors is

$$\mathbb{P}\bigl(m \text{ neigh. are 0-sand}, n \text{ neigh. are 1-sand}, (3-m-n) \text{ neigh. are 2-sand} \mid i\text{-sand node}\bigr) = \frac{3!(\phi_{i0})^m (\phi_{i1})^n (\phi_{i2})^{3-m-n}}{m!n!(3-m-n)!}, \tag{SM12}$$

while the values observed in $\mathcal{S}_{\text{BTW}}^\mu$ are slightly different from these predictions. Note that our general modeling approach allows for the future consideration of such higher-order correlations.

In the limit $\epsilon \to 0$ (obtained by linear extrapolation), $\mathcal{M}_{\text{BTW}}^\mu$ predicts $\phi_{00} \to 0$ and $\phi_{22} \to 1/2$. Both of these predictions can be understood intuitively. For $\epsilon \to 0$ and under the assumptions of $\mathcal{M}_{\text{BTW}}^\mu$, $\phi_{22} = 1/2$ is required to make the system critical (i.e., branching factor $R_0 = 1$). Moreover, a non-zero $\phi_{00}$ requires the presence of 00-sand links in the network, but 00-sand links become 01- or 10-sand links whenever one of the two nodes receives sand, and 00-sand links can only appear when a grain of sand dissipates along a 01- or 10-sand link, which occurs arbitrarily rarely for $\epsilon \to 0$. In fact, for $\epsilon > 0$, we see $\phi_{00} \approx \epsilon/2$.

We have no reason to offer as to why some other $\phi_{ij}$ appear to take the values of simple ratios. For example, for $\mathcal{M}_{\text{BTW}}^\mu$ in the limit $\epsilon \to 0$, we seem to have $\phi_{20} \to 1/8$ and $\phi_{21} \to 3/8$. Note that this may well be a coincidence.

## SM7. OPPORTUNITIES FOR CONTROLLING THE CASCADE AREA

Our $\mu$-control scheme attempts to mitigate costs associated with cascade size (the number of topplings). However, because sand input and dissipation must balance and because average cascade size determines average dissipation, the cascade size and hence our control scheme are constrained.

In situations like this one, it may be possible to control the cascade area much more than the cascade size. In other words, if many topplings must occur in order to dissipate enough sand, perhaps we can isolate those topplings so that the cascade area (the number of nodes that topple) is small. To do so, the controller could try to concentrate at-capacity nodes to a small region of the network (and somehow continue to concentrate at-capacity nodes the long run).

Here we bound the extent to which one may decrease the ratio $a/s$, where $a$ is the cascade area and $s$ is the cascade size. We provide arguments supporting the claim that the expansion of the network structure (i.e., the rate at which the number of nodes at distance $n$ scales with the number of nodes at distance at most $n$) severely limits how low $a/s$ may get. Specifically, in tree-like graphs, the cascade area can be as small as half the cascade size, whereas arbitrary small $a/s$ ratios may be possible in graphs with less expansion than tree-like graphs (e.g., networks with community structure, finite-dimensional lattices). Note that we do not provide control strategies to reach these bounds, nor do we assert that reasonable control strategies approaching these bounds exist.

We first substantiate the conjecture that $1/2 < a/s \leq 1$ for random 3-regular graphs in the infinite network limit (i.e., $N \to \infty$). These arguments could generalize to large tree-like graphs, such as the giant connected component of a configuration model random graph. Consider the following "situation †" that results in a cascade area $a^\dagger$ and a cascade size $s^\dagger$: the root is initially at capacity (2-sand), every node $< n$ hops away from the root is initially at capacity, each node exactly $n$ hops away from the root is not at capacity (0- or 1-sand), and no dissipation occurs during the resulting cascade. We will later see that, in some sense, $a^\dagger/s^\dagger = \inf a/s$ for an infinite random 3-regular graph, with the infimum taken over all possible cascades.

In situation †, Corollary 1 shows the following: the root topples $n$ times and has pattern (v) with all children of signature $(n, n-1)$; nodes exactly $1 \leq \ell < n$ hops away from the root topple $n-\ell$ times and have pattern (vi) with all children of signature $(n-\ell, n-\ell-1)$; nodes exactly $n$ hops away do not topple and have the pattern (ii); and nodes $> n$ hops away away do not topple and do not have a pattern. Hence, the area $a^\dagger$ is given by

$$a^\dagger = 1 + 3 \sum_{\ell=1}^{n-1} 2^{\ell-1} = 3(2^{n-1}) - 2, \tag{SM13}$$

whereas the cascade size $s^\dagger$ is

$$s^\dagger = n + 3 \sum_{\ell=1}^{n-1} 2^{\ell-1}(n-\ell) = 3(2^n) - 2n - 3. \tag{SM14}$$

Thus, the ratio of cascade area to size is

$$\frac{a^\dagger}{s^\dagger} = \frac{3(2^{n-1}) - 2}{3(2^n) - 2n - 3} \xrightarrow{n \to \infty} \frac{1}{2}. \tag{SM15}$$

Because $a^\dagger/s^\dagger$ converges to $1/2$ monotonically from above, in situation † the cascade area is at least half the cascade size.

We now provide support that, in a sense, situation † minimizes the ratio $a/s$. From Corollary 1, no node may topple more times than the root (which is $n$ times in situation †), and a node cannot topple more times than the length of the shortest path from this node to a node initially not at capacity (which is $n - \ell$ times in situation † for a node $\ell$ hops away from the root). Now consider a "situation ‡" that contains a "kernel" of nodes less than $m$ hops from the root that are all at capacity; we define $n$ as the largest $m$ satisfying this constraint, $a^\ddagger$ as the resulting cascade area and $s^\ddagger$ as the resulting cascade size. In the absence of dissipation, one may show that the ratio $a^\ddagger/s^\ddagger$ is minimal if nodes exactly $n$ hops away from the root are all initially not at capacity, which effectively gives situation †.

Finally, consider a situation where some dissipation occur in a cascade that began in "situation †". If the first grain sent from a parent node to its child dissipates, then by Corollary 1 [parts (iii) and (iv)] the child cannot topple during this cascade. The cascade size and area would have been the same as in the situation in which the grain does not dissipate if the child had been initially not at capacity, and we have already seen that this situation in which the child is not at capacity may not decrease the ratio $a/s$. If any other grain (i.e., a grain sent from a child to its parent, or an $m$-th grain with $m > 1$ sent by a parent node to its child) dissipates, then by Corollary 1 this dissipation cannot affect the area of the cascade, and it may not increase its size. Hence, dissipation may not decrease the ratio $a/s$. This completes the discussion about $1/2 < a/s \leq 1$ for large random 3-regular graphs.

Next we show that arbitrarily small ratios $a/s$ may be attainable in graphs with slower expansion. First consider the trivial example of two degree-1 nodes joined by a link. Both nodes have a capacity of 0, so if either of these nodes is the root of a cascade, then the two nodes will exchange the grain until it dissipates, hence typically resulting in a cascade area of $a = 2$ (assuming that the first grain shed does not dissipate, otherwise $a = 1$), while the cascade size is $1/\epsilon$ in expectation [because the cascade size is a geometrically-distributed random variable with parameter $\epsilon$; specifically, $\mathbb{P}(\text{size} = s) \equiv (1 - \delta_{0s})(1 - \epsilon)^{s-1}\epsilon$]. Hence, the ratio $a/s \approx 2\epsilon$ may clearly be less than $1/2$ for small enough $\epsilon$. A similar phenomenon occurs in a "tetrahedron graph" consisting of 4 degree-3 nodes that are all initially at capacity, or in any small graph consisting of nodes all at capacity.

Now consider the less trivial example of a network with community structure (i.e., with subgraphs that have many more connections within them than between them). Suppose that the root belongs to a community of $M$ nodes all at capacity, and all the nodes of this community are at least $m$ hops away from any node not at capacity. Denote by $L$ the number of nodes that are within $m-1$ hops of any node of this community but that are not part of the community. In the absence of dissipation, we have $a \geq M + L$ and $s \geq Mm + L$. If the community is large ($M$ large) and if the cascade is relatively confined to that community ($L$ small), then the ratio $a/s$ may reach values of the order of $1/m$ or even lower.

Finally consider the (non-random) 3-regular graph given by a hexagonal lattice. [For example, to define the graph on $\mathbb{Z}^2$, each pair $(x, y) \in \mathbb{Z}^2$ corresponds to a node that has the following 3 neighbors: $(x, y - 1)$, $(x, y + 1)$ and $(x + (-1)^y, y + (-1)^y)$.] Consider "situation $\Delta$", related to situation † above, in which all the nodes less than $n$ hops from the root are at capacity; the nodes exactly $n$ hops from the root are not at capacity; and no dissipation occurs. In this case, we may show that the cascade area is given by

$$a^\Delta = 1 + 3\sum_{\ell=1}^{n-1} \ell = 1 + \frac{3n(n-1)}{2};  \quad (SM16)$$

the cascade size is given by

$$s^\Delta = n + 3\sum_{\ell=1}^{n-1} \ell(n - \ell) = \frac{n(n^2 + 1)}{2}; \quad (SM17)$$

and their ratio is thus

$$\frac{a^\Delta}{s^\Delta} = \frac{3(n-1)}{(n^2+1)} \xrightarrow{n \to \infty} 0. \quad (SM18)$$

Future work could explore ways to control the BTW sandpile process $\mathcal{S}_{\text{BTW}}$ in order to mitigate cascade area in the equilibrium state of the system, which would presumably require knowledge of the distribution of sand over the nodes at each time step. Efficiently observing the state of the network at mesoscopic or macroscopic scales could use techniques for observing networks using few sensors [3, 4]. This approach of limiting the area of damage (provided that damage must occur) might even find applications in real infrastructure.


%
%****************************************************************
\end{document}